\documentclass[12pt,a4paper]{article}
\usepackage{amsmath}
\usepackage{amsthm}
\usepackage{amsfonts}
\usepackage{amssymb}
\usepackage{booktabs}
\usepackage{minitoc}
\usepackage{graphicx}
\usepackage{color}
\definecolor{darkgreen}{rgb}{0,0.5,0}
\definecolor{purple}{rgb}{1,0,1}
\usepackage{wrapfig}
\usepackage{pb-diagram}
\usepackage{latexsym}
\newcount\Comments  %
\usepackage{pdfpages}
\usepackage{xcolor}
\usepackage{mathtools}
\usepackage{enumitem}

\usepackage{wrapfig}
\usepackage{mdframed}
\usepackage{tikz}
\usetikzlibrary{angles}
\usetikzlibrary{patterns}
\usetikzlibrary{braids}
\usepackage{pgfplots}
\usepgfplotslibrary{fillbetween}
\usetikzlibrary{pgfplots.fillbetween} %
\usetikzlibrary[pgfplots.fillbetween] %

\newcommand{\kibitz}[2]{\ifnum\Comments=1\textcolor{#1}{#2}\fi}

\numberwithin{equation}{section}
\usepackage{srcltx}
\usepackage{authblk}
\usepackage[left=2.00cm, right=2.00cm, top=2.00cm, bottom=2.00cm]{geometry}
\theoremstyle{definition}

\newtheorem{exmp}{Example}[section]
\begin{document}
\title{Extended symmetry of higher Painlev\'e   equations of even
periodicity and their  rational solutions}

\author[1]{ H. Aratyn}
\author[2]{J.F. Gomes} 
\author[2]{G.V. Lobo} 
\author[2]{A.H. Zimerman}
\affil[1]{
Department of Physics, 
University of Illinois at Chicago, 
845 W. Taylor St.
Chicago, Illinois 60607-7059, USA}
\affil[2]{
Instituto de F\'{\i}sica Te\'{o}rica-UNESP,
Rua Dr Bento Teobaldo Ferraz 271, Bloco II,
01140-070 S\~{a}o Paulo, Brazil}

\maketitle

\abstract{ %
The structure of extended  affine Weyl symmetry group of 	 higher    Painleve equations of $N$ periodicity depends on whether $N$ is even or odd.
We find that for even 
$N$, the symmetry group ${\widehat A}^{(1)}_{N-1}$  contains the conventional B\"acklund
transformations $s_j, j=1,{\ldots},N$, the group of automorphisms
consisting of cycling permutations but also  
reflections on a periodic circle of $N$ points, which is a novel feature uncovered in this paper. 
 
	The presence of reflection automorphisms is connected  to  
	 existence of degenerated solutions and for $N=4$ we explicitly show  how the reflection 
	 automorphisms around even points cause  degeneracy of a class of
	 rational solutions obtained on the orbit of translation operators 
	 of ${\widehat A}^{(1)}_{3}$.  We obtain the closed expressions for solutions and
	 their degenerated counterparts 
	 in terms of determinants of Kummer polynomials.
	
  }
\section{Introduction}
\label{section:intro}

This paper analyses  the higher   even $N$ Painlev\'e  equations: %
\begin{equation} \begin{split} g_{i, \, x}&= 
\sum_{k=1}^{\frac{N}{2}-1} g_i g_{i+2k} \left(-\sum_{r=1}^{k} g_{i+2r-1}
+\sum_{s=k}^{\frac{N}{2}-1} g_{i+2s+1}\right)\\
&-\frac{\alpha_i+\alpha_{i+2}+\cdots +\alpha_{i+N-2}}{2x}
g_i+\frac{ \alpha_i}{2x},  \;\; i=1,{\ldots} ,N\, ,
\label{Nevengeneral}
\end{split}
\end{equation}
with $g_i$ and $\alpha_i$ satisfying  conditions
\begin{equation}
\sum_{r=1}^{N/2} g_{2r}=1, \quad 
\sum_{r=1}^{N/2} g_{2r+1}=1, \quad \sum_i^N \alpha_i=2\, ,
\label{vinculos}
\end{equation}
that are derived from the formalism of dressing chain equations 
\eqref{dressingeqseven} of even $N$ periodicity, meaning that
$g_{i+N}=g_i,\alpha_{i+N}=\alpha_i$, see also the formulation
of higher Painlev\'e equations in
\cite{NY1998} in terms of variable $z$ such that $x=-z^2/2$.

The  higher  Painlev\'e  equations \eqref{Nevengeneral}
are manifestly invariant under
the Weyl group of B\"acklund symmetry transformations $s_i,
i=1,2,{\ldots} ,N$ \eqref{backsymham} as well as the group of automorphisms
$\{ 1, \pi,{\ldots} , \pi^{N-1}, {\widehat P}_n, n=1,{\ldots} ,N \}$
where $\pi$ is a cyclic permutation defined in \eqref{backsymham} 
and reflection automorhisms ${\widehat P}_n$ are   
generated 
by reflections $P_n$ on a periodic circle of $N$ points \eqref{P2ik}
and \eqref{P2i1k}
and induce actions  of ${\widehat P}_{n}$
on $g_i(x), \alpha_i$ \eqref{representationoddeven} and  on
the B\"acklund symmetry transformations \eqref{representationsN}.

The reflections around even points $n=2i$ of the periodic circle
with the period $N$ are shown to 
produce degenerated solutions (solutions that share 
the same parameters $\alpha_i$ as original solutions), when acting on
solutions generated from the particular seed solution by translation
operators.

The paper is organized as follows.
After we  derive higher ${\widehat A}^{(1)}_{N-1} $ symmetric Painlev\'e  
equations 
from the dressing chain equations of even periodicity 
in section \ref{section:dress}, we analyze their symmetries
in section \ref{section:symPa}. 
In this section, we introduce 
reflection transformations $P_n, n=1,{\ldots} ,N$ that 
close under multiplication with cyclic permutations 
$\{ 1, \pi, \pi^2, {\ldots} , \pi^{N-1} \}$ of   the 
extended Weyl group ${\widehat A}^{(1)}_{N-1} $ and form with them
a group of automorphisms of higher  Painlev\'e  equations of even $N$
periodicity.  
We refer to these transformations as reflection
automorphisms and observe how they naturally split into even and odd
reflections depending whether the reflection is around even or odd point
of $1, {\ldots} , N$ periodic circle. 
All reflection automorphisms square to one and transform
the underlying variable $x$ to $-x$.
We describe representation of reflections  automorphisms
on the  $A^{(1)}_{N-1} $ symmetries $s_i, i=1,{\ldots} , N$ and
the abelian subgroup of $A^{(1)}_{N-1} $ consisting of 
translation operators $T_i, i=1,{\ldots} , N$.
These results path the way to an explanation  how the even
reflections introduce degeneracy among a class of rational solutions of 
 Painlev\'e  equations, which is  explicitly done for the $N=4$ model.

Working with the $N=4$ case
we present in section \ref{section:N4}  
an explicit construction of a class of %
rational solutions of Painlev\'e equations that reside 
on the orbit of translation 
operators of the extended affine Weyl group $A^{(1)}_3$.
These solutions are given by two components 
$F^{k}_{m} (x,\mathsf{a}), \, G^{k}_{m} (x,\mathsf{a})$ and   are labeled by two 
independent positive integer parameters that enter the formalism as
powers of the translation operators. We
find an explicit determinant expressions for $F^{k}_{m} (x,\mathsf{a}), \, G^{k}_{m} (x,\mathsf{a})$
in terms of Kummer polynomials in section \ref{section:Kummer}
for all orbit solutions generated from a particular seed solution.
The technique we use is based on the fact that the actions of translation operators 
give rise to the recurrence relations. We are able to 
solve these recurrence relations in terms of determinants of 
Kummer polynomials constructed
explicitly in this section. %
It is crucial to emphasize the role
played by one of the odd reflection automorphism that gives rise 
to the ``duality'' relation that ensures the intrinsic consistency of
rational solutions by permiting two dual formulations based on two related
Kummer polynomial definitions. 

In section \ref{section:dege}, we discuss the degeneracy that exists 
for a class of rational solutions of $N=4$ linking it to actions
of the two even reflection automorphisms that are shown to map the
rational solutions to different solutions that share the values of the
parameters of the $N=4$ Painlev\'e equations. 

The two remaining odd reflection automorphisms are discussed in section 
\ref{section:oddref}. The ${\widehat P}_1$  automorphism 
is shown to connect  two dual solutions. The invariance under
${\widehat P}_1$ ensures validity the recursive relations 
and accordingly proves the intrinsic consistency of construction 
 of the determinant solutions  based on these recursive relations.

In section \ref{section:higherN}, we show examples of how our formalism
extends to $N=6,8$ by utilizing presence of reflection automorphisms 
as a tool to finding a class of relevant solutions.

In appendix \ref{appendix:matrices} we illustrate some of our findings
by explicitly constucting braids and matrix representations of
reflection automorphisms for $N=4$.

\section{Dressing chain derivation of Painlev\'e equations for even
periodicity}
\label{section:dress}
We start with  a formalism of the  dressing chain equations of even periodicity.
The conventional definition of a dressing chain of $N$ periodicity is
\begin{equation}
(j_n+j_{n+1} )_z = -j_n^2+j_{n+1}^2 +\alpha_n, \;\; n=1,{\ldots} , N,
\quad j_{N+i}=j_i\, ,
\label{dressingeqs}
\end{equation}
however for even $N$   this expression requires, for consistency,  
an imposition of a  quadratic constraint that can be introduced as 
 modification of the dressing chain
formulation. Such modification was proposed on basis of Dirac reduction
method in \cite{AGZ2021}, where the authors put forward  
a system of dressing chain  equations 
of even periodicity  defined as :
\begin{equation}
(j_i+j_{i+1} )_z = -j_i^2+j_{i+1}^2 +\alpha_i
+(-1)^{i+1} \frac{(j_i+j_{i+1}) }{\Phi}\Psi
, \;\, i=1,2, {\ldots}
, N, \quad j_{N+i}=j_i\, ,
\label{dressingeqseven}
\end{equation}
where 
\begin{equation}
 \Psi=  \sum_{k=1}^N (-1)^{k+1} \left( j_k^2
- \frac12  \alpha_k \right), \qquad \Phi=\sum_{k=1}^N j_k=z \, .
\label{psidef}
\end{equation}
Remarkably, the dressing chain equations \eqref{dressingeqseven} can be rewritten
entirely in terms of $f_i=j_i+j_{i+1}$ without any references 
to $j_i$. %
It needs to be
emphasized that such elimination of $j_i$ variables while expressing
the dressing chain equations in terms of $f_i$ requires inserting the
definition of $\Psi$ from \eqref{psidef} into equations \eqref{dressingeqseven}.
This is because the relation $f_i=j_i+j_{i+1}$ is not invertible
for even  $N$. In contrast when $N$ is taken to be odd
such relation is invertible as illustrated by example of  $N=3$ with the
inverse relation given by:
\[ j_1 =\frac12 ( f_1-f_2+f_3), \;\;  j_2 = \frac12 ( f_2-f_3+f_1), \;
\;
j_3  = \frac12 ( f_3-f_1+f_2)\,,
\]
and after inserting these relations back  into equations \eqref{dressingeqs}
one obtains the symmetric Painlev\'e IV equations $f_{i,z}=
 f_i (f_{i+1}-f_{i-1})+\alpha_i, i=1,2,3$.

In the case of equations \eqref{dressingeqseven} 
introducing  new variables 
\begin{equation} 
\label{defx}
g_i(x) =\frac{f_i}{z}= \frac{j_i+j_{i+1}}{z},\quad x=-z^2/2\, ,
\end{equation}
we are able to derive 
the $N$ even  Painlev\'e equations \eqref{Nevengeneral},
where $g_i$ and $\alpha_i$ satisfy  conditions \eqref{vinculos}.
The equations \eqref{Nevengeneral} 
agree with higher Noumi-Yamada equations
\cite{NY1998} with $A^{(1)}_l$ symmetry for $l=2n+1$,
which were originally written  
in terms of variable $z$. %
One of the advantages of using $x$ is that
it makes it possible to uncover 
a new symmetry structure of $N$ automorphisms. These automorphisms
preserve the  $A^{(1)}_{N-1}$ Weyl symetry group of B\"acklund
transformations of equations \eqref{Nevengeneral} and will explain
presence of denegeneracy among rational solutions. 
In case of equations with  $N=4$ we obtain a regular 
Painlev\'e V equation 
directly from \eqref{Nevengeneral} as Painlev\'e V equation
is conventionally expressed using the variable $x$ (see below the equation
\eqref{xhameqs} for the derivation).

\section{Symmetries of higher Painlev\'e equations for even $N$} 
\label{section:symPa}
Equations \eqref{Nevengeneral} are invariant under the 
extended affine $A^{(1)}_{N-1}$ Weyl group of
B\"acklund transformations, $s_i$ and an automorphism $\pi$ : %
\begin{equation}
 \begin{split}
s_i &: g_{i \pm 1} \to g_{i \pm 1} \mp \frac{\gamma_i}{g_i} \,,\;\; \;
g_{i } \to g_{i },\;\; \; g_{j } \to g_{j }, \,j \ne i, i \pm 1 ,\\
& \gamma_i \to -\gamma_i, \; \; \gamma_{i \pm 1} \to \gamma_{i \pm 1}
+\gamma_i , \; \;  \gamma_{j} \to  \gamma_{j},\, j \ne i, i \pm 1, \;\;
i=1,{\ldots}, N\, ,\\
\pi  &:  \pi(g_i)= g_{i+1} ,\;  \pi(\gamma_i)=  \gamma_{i+1},
 \label{backsymham}
 \end{split}
 \end{equation}
where for convenience we introduced the parameters
\[ \gamma_i=\frac{\alpha_i}{2x} \]
with the normalization $\sum_i^N  \gamma_i=1/x$ that corresponds to
normalization $\sum_i^N  \alpha_i=2$ for the parameters $\alpha_i$.

In addition to the 
extended affine Weyl group of B\"acklund transformations
the equations \eqref{Nevengeneral} are invariant under
automorphisms that %
are generated by reflections around the fixed point $n$ on a periodic circle $1,2,,3,{\ldots} ,N+1,
N+2,{\ldots} $ with a periodic condition $k=k+N$ for any point $k$. 
These automorphisms also map $x \to- x$. 
We denote these automorphisms as $P^{N}_n$ but will drop the
upper script $N$, 
when it will be obvious from the
context and will use only the $P_n$ symbol. %
We distinguish between
two types of automorphisms 
depending on whether
$n$ is even or odd.  Accordingly, the two classes of reflections are  :
\begin{enumerate}
\item Reflections around even points $n=2i$ denoted by 
$P_{2i}, i=0,1,{\ldots} , \frac{N}{2}$ 
that act  as follows: 
\begin{equation}
P_{2i} (k)=2i-k,\quad k=0,1,2,3,4,5,..,\quad k+N=k 
\label{P2ik}
\end{equation}
on a periodic line of points  with a period of $N$.
Each reflection is equivalent to 
a single transposition  that interchanges
 $k$ with $2i -k$.
There are always two fixed points, namely,  
 $i$ and $i+\frac{N}{2}$ that according to
\[
P_{2i} (i)= 2i-i=i, \;\;\; P_{2i} (i+\frac{N}{2})= 
2i-i-\frac{N}{2}=i-\frac{N}{2} \sim i+\frac{N}{2} ,
\]
that are being transformed into themselves. The symbol $\sim$ is
to indicate that the periodic condition $k=k+N$ was used.
The remaining $N-2$ points %
fall into 
$N/2-1$ (even/even and odd/odd) pairs that are mapped into each other
under even reflections.

\item Reflections around odd points $n=2i+1$ denoted by 
$P_{2i+1}, i=0,1,{\ldots} , \frac{N}{2}-1$ 
that act  through 
\begin{equation}
P_{2i+1} (k)=2i+1-k 
,\quad k=0,1,2,3,4,5,..,\quad k+N=k 
\label{P2i1k}
\end{equation}
on a periodic line of points $k=1,2,3,4, {\ldots} $ with a period of $N$.
Each of  these reflections is  equivalent to transpositions
of $N/2$ pairs of even/odd points that are being interchanged
into each other.
\end{enumerate}
All these reflections obviously square to one: 
\[P_n^2=1 \,.\]

All the above  reflections  naturally induce transformations of $\alpha_k, g_k$
through:
\begin{equation}
\begin{split}
{\widehat P}_{n} (g_k (x))&=g_{P_{n} (k)} (-x)= g_{n-k}(-x) , \\
{\widehat P}_{n} (\alpha_k)&=\alpha_{P_{n} (k)} = \alpha_{n -k} , \;\; n=2i,
2i+1
\end{split}
\label{representationoddeven}
\end{equation}
that keep equations \eqref{Nevengeneral} invariant when also $x \to -x$.
The straightforward although formal way to prove the invariance of 
the Painlev\'e
equations \eqref{Nevengeneral} is to introduce transformations :
\begin{equation} {\widehat P}_{n} (j_k)= \sqrt{-1} j_{n-k+1}, \;\; {\widehat P}_{n} 
(\alpha_k)= \alpha_{n -k}, \;\; {\widehat P}_{n} (z) = \sqrt{-1} z \,.
\label{pnjk}
\end{equation}
of the dressing equations \eqref{dressingeqseven} 
that generate transformations \eqref{representationoddeven} together
with $x \to-x$. It follows from these definitions that
\[{\widehat P}_{n} (\Psi)= (-1)^n \Psi, \quad 
{\widehat P}_{n} (\Phi)= \Phi,
\]
when the use is made of periodic conditions.
As a result the dressing chain equations \eqref{dressingeqseven}
remain invariant under the transformations \eqref{pnjk}. This in
turn proves invariance of equations \eqref{Nevengeneral} under 
automorphisms \eqref{representationoddeven} that %
are generated by reflections on a periodic circle.
Note that the imaginary factor  present in definition \eqref{pnjk}
goes away when the transformation is applied on quantities 
$x$ and $g_k (x)$.

We find from the definitions of reflections  the following relations
for obtaining the automorphisms $\pi$ and its powers out of products of 
reflections ${\widehat P}_n $:
\begin{equation}
 \pi = {\widehat P}_{n+1} {\widehat P}_n , \; 
\pi^{-1}= {\widehat P}_n {\widehat P}_{n+1}, \, 
 \pi^2= {\widehat P}_{n+2} {\widehat P}_n , \; 
\pi^{-2}= {\widehat P}_n {\widehat P}_{n+2},\;\; n=2 i,2 i+1\, .
 \label{piRn}
 \end{equation}
More generally we find the following identities
\begin{equation}
{\widehat P}_{n_1}{\widehat P}_{n_2}= \pi^{n_1-n_2},\quad
{\widehat P}_{n_1} \pi^{n_2}={\widehat P}_{n_1-n_2}\, ,
\label{groupauto}
\end{equation}
which show that reflections and $\pi$ automorphisms close under
multiplications and form the group with identity $1=\pi^{0}=
{\widehat P}_{n}^2$. This structure is consistent with the reflection
automorphisms transforming $x \to - x$, despite the fact that $\pi$ 
is not acting on $x$.
We will refer to reflections and  $\pi$ and its powers 
 as group of
automorphisms with cyclic permutations 
$\{ 1, \pi, \pi^2, {\ldots} , \pi^{N-1} \}$  forming a subgroup of these
automorphisms.

Even/odd reflections commute among themselves
 \[ \lbrack P_{2i},P_{2j}  \rbrack =0 ,  \;\quad
 \lbrack P_{2i+1},P_{2j+1} \rbrack =0 , \;\; i \ne j
 \]
while we have the following conjugation transformations:
 \begin{equation}\begin{split} P_{2i+1} P_{2i} P_{2i+1}
&= P_{2i+2} \;\;,\;\; P_{2i+1} P_{2i} P_{2i+3}
= P_{2i} \\
P_{2i} P_{2i+1} P_{2i}
&= P_{2i+1} \;\;,\;\; P_{2i} P_{2i+1} P_{2i+2}
= P_{2i-1} 
 \end{split}
 \label{conjugation}
 \end{equation}
for the mixed even/odd reflections.

Additional important identities are :
\begin{equation}
{\widehat P}_n  \pi {\widehat P}_n = \pi^{-1},\;\;\; {\widehat P}_n  \pi^{-1} 
{\widehat P}_n = \pi,\; 
n=2i,2i+1\, .
\label{pim1pi}
\end{equation}
It is important that reflections preserve the B\"acklund symmetries
through the following conjugation formula:
\begin{equation}
{\widehat P}_n  s_m {\widehat P}_n \,= \,s_{P_n(m)},\; 
n=2i,2i+1,\, n, m=1,{\ldots} ,N
\label{representationsN}
\end{equation}
where $P_n(m)$ is the point obtained from $m$
by acting with the reflection $P_n$.

The above situation differs from the one which is encountered in
the case of odd $N$. First, in such case the corresponding 
Painlevé equations remain expressed in terms of the variable $z$ that
causes a presence of imaginary factors when dealing with
the transformation defined in \eqref{pnjk}.
Furthermore the ``even'' and ``odd'' reflections are no longer
clearly distinguishable. 
For example, for $N=3$ trying to apply the above definitions of reflections will
result in a similar action for the ``even'' reflection:
\[ P_2 (1)=1,\;  P_2 (2)=3,\; P_2(3) =2\, ,\]
as for the ``odd'' reflections:
\[ P_1 (1)=3,\;  P_2 (2)=2,\; P_2(3) =1, \; 
P_3 (1)=1,\;  P_3 (2)=1,\; P_3(3) =3\, .\]
We notice in all three cases the existence of 
one fixed point with the remaining two points being
transposed into each other. Although such reflections are not inducing 
directly any symmetry of the
symmetric Painlev\'e IV equations (without introducing a transformation
of the variable $z$ involving imaginary factor), a similar but 
albeit more general automorphisms have
been recently discussed in \cite{victor}.

\section{A special  case of $N=4$ and Painlev\'e V equations}
\label{section:N4}
For $N=4$ the equations \eqref{Nevengeneral} 
simplify to :
\begin{equation}
\begin{split}
g_{i\,,x}&=g_i g_{i+2} (g_{i-1}-g_{i+1}) -
\frac{\alpha_i+\alpha_{i+2}}{2x} g_{i}+\frac{\alpha_i}{2x}\\
&=g_i g_{i+2} (g_{i-1}-g_{i+1}) +\frac{\alpha_i}{2x} g_{i+2}
-\frac{\alpha_{i+2}}{2x} g_{i}\, , \;\; i=1,2,3,4
\end{split}
\label{symFG}
\end{equation}
We impose  everywhere in this section the conventional periodicity
conditions : $g_{i}=g_{i+4}$, $\alpha_{i}=\alpha_{i+4}$.

We write $g_1=F,g_2=G$, $g_3=1-F,g_4=1-G$
and $\alpha_4=2-\sum_{i=1}^3 \alpha_i$, with $F,G$ satisfying equations:
\begin{equation}\begin{split}
F_x &=F(F-1)(2G-1)-\frac{\alpha_1+\alpha_3}{2x} F +
\frac{\alpha_1}{2x} \, ,\\
G_x &= -G (G-1)(2F-1) + \frac{\alpha_1+\alpha_3-2}{2x} G
+\frac{\alpha_2}{2x} \, ,
\label{xhameqs}
\end{split}
\end{equation}
derived from equations \eqref{symFG}. In this setting the $A^{(1)}_{3}$ extended affine Weyl group
of  B\"acklund transformations, $\pi, s_i,i=1,{\ldots}, 4$
emerges as group of symmetry operations on equations \eqref{xhameqs} 
 and can be 
obtained from relations \eqref{backsymham} for $N=4$.

Below we will determine
explicitly solutions
of equations \eqref{xhameqs}
for the special values of the parameters $\alpha_1,\alpha_2,
\alpha_3$:
\begin{equation}
 \alpha_1=\mathsf{a} + 2k,\; \; \alpha_2= -2 k, \; \; \alpha_3=-2m,
\; \;  m,k \in \mathbb{Z}_{+}\, ,
\label{alphavalues}
\end{equation}
where $ \mathbb{Z}_{+}$ contains positive integers and zero.
In the notation of four-component parameters $\alpha_i, i=1,2,3,4$ we
have \[
\alpha_i= 2\, (\frac{\mathsf{a}}{2}+k, -k, -m , 1-
\frac{\mathsf{a}}{2}+m)\, ,
\]
with $\mathsf{a} $ being an arbitrary variable.
For the values of parameters given in \eqref{alphavalues} the solutions
of equations \eqref{xhameqs} can be obtained by action of translation
operators on a class of seed solutions as will be presented
in the next few subsections. %

Let us finally comment on connection of equations \eqref{xhameqs} 
to the Painlev\'e V equation 
that can be recovered   by eliminating
one of the two variables. 
Eliminating $G$ from equations \eqref{xhameqs} we find that 
$y=F (F-1)^{-1}$
satisfies the Painlev\'e V equation:
\[ 
y_{x x}
= -\frac{y_x}{x}+\left( \frac{1}{2y}+ \frac{1}{y-1}\right)
y_{x}^2 + \frac{(y-1)^2}{x^2} \left({ \alpha} y + 
\frac{{ \beta}}{y} \right) + \frac{{ \gamma}}{ x} y + { \delta} 
\frac{y (y+1)}{y-1}\, .
\]
with 
\[{ \alpha}= \frac18 \alpha_3^2,\;\; { \beta}=- \frac18 \alpha_1^2,\;\;
{ \gamma}=  \frac{\alpha_4-\alpha_2}{2  }, \, \delta= - \frac12
\]
\[
\alpha=  \frac12  \left( m\right)^2, \;\;
{ \beta}= - \frac12 \left( \frac{\mathsf{a}}{2}+k\right)^2,  \;\;
\gamma=1-\frac{\mathsf{a}}{2}+m+k, \, \;  m,k \in \mathbb{Z}_{+}\, .
\]
Eliminating instead $F$ from equations \eqref{xhameqs} we find that
$y=G (G-1)^{-1}$ 
satisfies the Painlev\'e V equation with  coefficients:
\[{ \alpha}= \frac18 \alpha_4^2,\;\; { \beta}=- \frac18 \alpha_2^2,\;\;
{ \gamma}=  \frac{\alpha_1-\alpha_3}{2  }, \, \delta= - \frac12
\]

\subsection{Seed solutions and translation operators}
\label{subsection:seedtrans}
There are essentially two fundamental seed solutions to equations
\eqref{xhameqs}:
\begin{align}
F&=\frac12, \quad G=\frac12, \quad
(\mathsf{a}, 1-\mathsf{a}, \mathsf{a}, 1-\mathsf{a})\, ,
\label{wata3}\\
F&=1, \quad G=0, \quad 
(\mathsf{a}, 0,0, 2-\mathsf{a}) 
\label{wata1}
\end{align}
with an arbitrary constant  $\mathsf{a}$. 
The first of these seed solutions given in \eqref{wata3} is invariant under 
$\pi^2$. This class of seed solutions gives rise to Umemura polynomials and
is not associated with degeneracy. 
It is illustrative to recall \cite{AGLZ2023}, that in the setting of
the even dressing chain \eqref{dressingeqseven}
the seed solution \eqref{wata3} is represented by 
solution $j_i (z)=(z/4)(1,1,1,1)$ 
with only positive components, while the seed solution \eqref{wata1} 
is represented by solution $j_i (z)=(z/4)(1,1,-1,1)$ 
with one negative and three positive components.
In this presentation, we will only study solutions \eqref{wata1}
to obtain closed expressions for
the special function solutions 
 generated generated from it by B\"acklund transformations.
Acting with $ \pi$ will generate from the solution \eqref{wata1}
equivalent solutions:  %
\begin{align}
 F&=1, \quad G=1, \quad( 2-\mathsf{a},\mathsf{a},0,0)\, , %
 \label{wata2}\\
 F&=0, \quad  G=1,\quad   (0, 2-\mathsf{a},\mathsf{a},0)  \label{wata5} \, ,\\
 F&=0,\quad G=0, \quad  (0,0, 2-\mathsf{a},\mathsf{a},
 )  \label{wata6} \, ,
 \end{align}
which therefore do not require a separate treatment.
We will refer to the seed solution \eqref{wata1}  using the following
(Physics inspired) notation:
\begin{equation}
\vert F=1,\, G=0  \rangle_{(\mathsf{a}, 0,0, 2-\mathsf{a}) } \, .
\label{vacuum}
\end{equation}

Within the $A^{(1)}_{3}$
extended affine Weyl group one defines an  abelian 
subgroup of translation operators defined as 
$T_i=r_{i+3} r_{i+2} r_{i+1} r_i, i=1,2,3,4$, 
where $r_i=r_{4+i}=s_i$ for $i=1,2,3$ and $r_4 = \pi$.
The translation operators commute among themselves, 
$ T_i T_j =T_j T_i$, and 
generate the following translations 
when acting on the $\alpha_i$ parameters:
\[ T_i (\alpha_i) = \alpha_i+2,\;  T_i (\alpha_{i-1})=\alpha_{i-1}-2,\;
 T_i (\alpha_j) =  \alpha_j, \; j=i+1, j=i+2\,.\]
The inverse translation operators are defined as $T^{-1}_i=r^{-1}_{i} r^{-1}_{i+1} 
r^{-1}_{i+2} r^{-1}_{i+3}$, with $r^{-1}_i=r^{-1}_{4+i}=s_i$ for $i=1,2,3$ 
and $r^{-1}_4 = \pi^{-1}$. They satisfy $T^{-1}_i T_i=1$ and
accordingly they  generate the following  $\alpha_i$ shifts
\[ T_i^{-1} (\alpha_i) = \alpha_i-2,\;  T_i^{-1} (\alpha_{i-1})=\alpha_{i-1}+2,\;
 T_i^{-1} (\alpha_j) =  \alpha_j, \; j=i+1, j=i+2\,.\]

We notice that the B\"acklund transformations $s_2,s_3$ generate
infinities  when applied
on the solution \eqref{wata1} because of $G=0$ and $1-F=0$.
To avoid these singularities %
only the following 
solutions are permitted to be generated 
out of the the seed solution \eqref{wata1} by use of
translations \cite{AGLZ2023}:
\[ 
T_1^{n} T_2^{-k} T_4^{m} \vert F=1,\, G=0  \rangle_{\alpha_{\mathsf{a}}}, \; n
\in \mathbb{Z},  \; k ,m \in \mathbb{Z}_{+}, \,
\alpha_{\mathsf{a}} = (\mathsf{a}, 0, 0, 2-\mathsf{a})\, .
\]
In this way we obtain, by action of translation operators,
new solutions
$ T_1^{n} T_2^{-k} T_4^{m} (F=1)$ and 
$ T_1^{n} T_2^{-k} T_4^{m} (G=0)$
of equations \eqref{xhameqs}
with a new
parameter:
\begin{equation}  T_1^{n} T_2^{-k} T_4^{m} (\alpha_{\mathsf{a}} )
=(\mathsf{a} +2 n+2 k,\, -2 k,\,
 -2 m,\,  2-\mathsf{a}-2 n +2 m)
\,.
\label{alphaT1T2T4}
\end{equation} 
One notices that the action of $T_1^{n} $ merely produces a shift of a 
parameter   $\mathsf{a}$ and as shown in \cite{AGLZ2023} leaves  the 
configuration $F=1, G=0$ unchanged:
\begin{equation}
T_1^{n} \vert F=1,\, G=0  
\rangle_{\alpha_{\mathsf{a}}}= \vert F=1,\, G=0  
\rangle_{\alpha_{\mathsf{a}+2 n}}.
\label{T1shift}
\end{equation}
We can therefore, 
replace the action of $T_1$ by appropriately redefining $\mathsf{a}$ 
and accordingly restrict our discussion to the solutions
of $F,G$ coupled equations of the form :
\begin{equation} \begin{split}
&\mathbb{T} ( k,m ; \mathsf{a})=  T_2^{-k} T_4^{m} \vert F=1,\, G=0  
\rangle_{\alpha_{\mathsf{a}}},  \quad \; k,m \in \mathbb{Z}_{+} \, ,\\
&\alpha_{k,m, \mathsf{a}}=  T_2^{-k} T_4^{m} (\alpha_{\mathsf{a}} )
=(\mathsf{a} +2 k,\, -2 m,\,
 -2 m,\,  2-\mathsf{a} +2 m)\, ,
\end{split}
\label{orbit}
\end{equation}
where we listed both the solution and its corresponding  parameter  $
\alpha_{k,m, \mathsf{a}}$ generated by combined actions
of translation operators
$T_2^{-k}$ and $T_4^{m}$, with $k,m$ being positive
integers, acting on the seed solution \eqref{wata1}.
In section \ref{section:Kummer}
we will find closed expressions for solutions $\mathbb{T} ( k,m ;
\mathsf{a})$ introduced in relation \eqref{orbit}
in terms of Kummer polynomials.

\subsection{Action of automorphisms for the $N=4$ model}
\label{subsection:autoN4}
Taking a look at the $N=4$ equations  \eqref{symFG}
we can easily see that they are manifestly  invariant %
under actions generated by four (two even and two odd)
automorphisms $P_i,i=1,2,3,4$ :
\begin{align}
P_4 &\to  : x \to  -x , \; g_1 \leftrightarrow g_3
,\;  \; \alpha_1 \leftrightarrow 
\alpha_3   \label{xmxF1mF}\\
P_2 &\to   : x \to -x, \; g_2  \leftrightarrow g_4,\; 
\alpha_2 \leftrightarrow 
\alpha_4   \label{xmxG1mG}\\
P_1 &\to  :x \to -x , \; g_1 \leftrightarrow g_4,\;
g_2\leftrightarrow g_3, 
\; \alpha_1 \leftrightarrow 
\alpha_4,\; \alpha_2 \leftrightarrow 
\alpha_3
\label{rho3prod}\\
P_3 &\to  :x \to -x , \; g_1 \leftrightarrow g_2,\; 
g_3\leftrightarrow g_4, \; \alpha_1 \leftrightarrow 
\alpha_2,\; \alpha_3 \leftrightarrow 
\alpha_4,   \label{xmxFG}
\end{align}
The above automorphisms are associated with 
reflections on a circle $k=1,2,3,4$ with periodic conditions
$k=5=1,k=6=2,{\ldots}$
obtained when we add periodic condition  $k+4=k$
to the line of points $k=1,2,3,4,5,..$ introduced around equations
\eqref{P2ik} and \eqref{P2i1k} for a general even $N$.
As in the case of general even $N$ there is an even number of
even and two odd automorphisms for $N=4$. 
Namely, there are even reflections $P_{2} , P_{4}$ introduced in relations 
\eqref{P2ik} that act as simple transpositions of two points.
The automorphisms  $P_{1} , P_{3}$ are odd reflections
introduced in \eqref{P2i1k}.
They all  act  on $\alpha_k, g_k$ via relations 
\eqref{representationoddeven}. 
\begin{exmp}
It follows that for $i=1$ the reflection 
${P}_{2i+1} ={ P}_{3}$ acts as 
\[
P_{3} (1) =2,\;\; P_{3} (2) =1,\;\; P_{3} (3) =0 = 4 ,
\;\; P_{3} (4) = -1 = 3
\]
and induces automorphism \eqref{xmxFG}.

Similarly for $i=2$ the reflection ${P}_{2i+1} ={P}_{5} \sim P_{1}$ acts
as 
\[
P_{5} (1) =4,\;\; P_{5} (2) =3,\;\;P_{5} (3) =2 ,
\;\; P_{5} (4) = 1 \, .
\]
and induces the automorphism \eqref{rho3prod}.
For even reflections it appears that they are effectively transpositions
of two lines. For example for $P_{2i}$ with $i=1$ we have using the
periodicity $i+N \sim i$ : 
\[
P_{2} (1) =1,\;\; P_{2} (2) =0  \sim 4,\;\;P_{2} (3) =-1 \sim 3,
\;\; P_{2} (4) = -2 \sim 2\, .
\]
Thus $P_{2}$ is transposing points $2$ and $4$ while $1$ and $3$ are
fixed points and consequently induces ${\widehat P}_{2}$
from relation \eqref{xmxG1mG}.
Similarly for $P_{2i}$ with $i=2$ we have
\[
P_{4} (1) =3,\;\; P_{4} (2) =2,\;\;P_{4} (3) =1 ,
\; \;P_{4} (4) = 0 \sim 4\, .
\]
Correspondingly  ${\widehat P}_{4} $ acts as shown in relation \eqref{xmxF1mF}. 
\end{exmp}

Below are few obvious identities following from the basic definitions:
\[ 
 P_{2}P_{3}= P_{4} P_{1} ,\;\; P_{4}P_{3}= P_{2} P_{5}
\]
or more generally
\[
P_{2i}P_{2i+1}= P_{2i+2} P_{2i+3}
\]
etc.
Also it follows that
\[
{\widehat P}_{2i+1} {\widehat P}_2 {\widehat P}_{2i+1}= {\widehat P}_{4},
\;  {\widehat P}_{2i} {\widehat P}_1 {\widehat P}_{2i}= {\widehat P}_{3},\]
etc. We refer to Appendix \ref{appendix:matrices} for a convenient 
matrix representation of such identities.

Furthermore, the product of all four reflections ${\widehat P}_4{\widehat P}_2 {\widehat P}_1 {\widehat P}_3$ is an
identity due to
\[ P_4 P_2 P_1 P_3(k)=-(2-(1-(3-k))=k-4=k\, .
\]

\subsection{Conjugations of B\"acklund  and translations operators with  reflections}
\label{subsection:conjugation}
Conjugation of B\"acklund transformations by
reflections is described by the general formula \eqref{representationsN}
adopted to $N=4$ case:
\begin{equation}
{\widehat P}_n  s_m {\widehat P}_n \,= \,s_{P_n(m)},\; 
n=2i,2i+1,\, n, m=1,2,3,4
\label{representations}
\end{equation}
where $P_n(m)$ is the point obtained from $m$
by acting with the reflection $P_n$.
Repeated conjugation yields:
\[{\widehat P}_{n+1} {\widehat P}_n s_m {\widehat P}_n {\widehat P}_{n+1} 
=s_{P_{n+1}P_n (m)}=s_{m+1}
\]
which in view of relations \eqref{piRn} reproduces a basic relation
$\pi s_m=s_{m+1}\pi$.

We are able to extend the above results
to include conjugations by  reflections of the 
translation operators.
They are described by the following two relations : %
\begin{equation}
	{\widehat P}_n T_i^{-1} {\widehat P}_n= T_{P_{n+1} (i)},\; \;
	{\widehat P}_n T_i {\widehat P}_n= T_{P_{n+1} (i)}^{-1} \, .
	\label{Rntranslation}
\end{equation}
Several relevant for $N=4$ examples that follow from the above identity
are listed bellow: %
\begin{align}
	{\widehat P}_2 T_2^{-1} {\widehat P}_2&= T_1,
	\; \;{\widehat P}_2 T_4 {\widehat P}_2= T_3^{-1} \, ,
	\label{R2translation} \\
	{\widehat P}_0 T_2^{-1} {\widehat P}_0&= T_3,\; \;
	{\widehat P}_0 T_4 {\widehat P}_0= T_1^{-1}\,.
	\label{R0translation}
\end{align}
The transformations of translation operators by odd reflections are as 
follows :
\begin{equation}
{\widehat P}_1 T_2^{-1}{\widehat P}_1= T_4,\quad {\widehat P}_1 T_4 {\widehat P}_1= T_2^{-1}\, ,
\label{R1translation}
\end{equation}
for ${\widehat P}_1$ reflection and 
\begin{equation}
{\widehat P}_3 T_2^{-1}{\widehat P}_3= T_2,\quad {\widehat P}_3 T_4 {\widehat P}_3= T_4^{-1}\, ,
\label{R3translation}
\end{equation}
for ${\widehat P}_3$ reflection. 
The first of these results given in relation
\eqref{R1translation}
is the duality map: $k \leftrightarrow m$  
that also maps $x \to-x$ and
$ \mathsf{a} \to 2-\mathsf{a}$, since $\alpha_1 \leftrightarrow \alpha_4$.

\section{ Closed
expressions for solutions on the orbit of translation operators.}
\label{section:Kummer}
In this section we find a closed  expression for the general solution
$\mathbb{T} ( k,m ;
\mathsf{a})$ introduced in relation \eqref{orbit}.
We start by  first considering  the two simpler cases of
$T_4$ orbit and the $T_2$ orbits.
\subsection{The $T_4$ and the $T_2$ 
orbits and Kummer polynomials}
\label{subsection:T2T4orbits}
The $T_4$ orbit and the $T_2$ 
orbits included in the definition of 
\eqref {orbit} are described as :
\begin{enumerate}
\item the $k=0$ case that maintains $G=0$, while it
transforms $F$ by $T_4$, 
\item the $m=0$ case that maintains $F=1$  and 
transforms $G$ by $T_2$.
\end{enumerate}
The $T_4$ orbit:
\[
  \mathbb{T} ( 0,m; \mathsf{a})=   T_4^{m} \vert F=1,\, G=0  
 \rangle_{\alpha_{\mathsf{a}}} = \big( F_m (x, \mathsf{a})\,, G_m(x, \mathsf{a})=0
 \big) \, ,
\]
is governed by the recurrence relation 
\begin{equation}F_{m} =1 +\frac{x m}{x F_{m-1}+\mathsf{a}/2-m}\, ,
\label{recT4}
\end{equation}
obtained from  the $T_4$ transformation rule
\begin{equation}
\begin{split}
T_4(F)&= 1-G+(\gamma_1+\gamma_4)/(F-\gamma_4/(1-G))\, , \\
T_4 (G) &=
F-\gamma_4/(1-G)+(\gamma_1+\gamma_2+\gamma_4)/(G-(\gamma_1+\gamma_4)/(F-
\gamma_4/(1-G)))
\, .
\label{T4pq}
\end{split}
\end{equation}
The above recurrence relation for $F_m$ 
has a solution :
\begin{equation} F_m= \frac{N_m(x,\mathsf{a}+2)}{N_m(x,\mathsf{a})}\, , \label{Fk0}
\end{equation}
in terms of Kummer polynomials $N_m(x,\mathsf{a})$ :
\begin{equation}
N_n (x,\mathsf{a}+2)=2^n x^n (-x)^{-\mathsf{a}/2} U(-\frac{\mathsf{a}}{2},-\frac{\mathsf{a}}{2}+n+1,x) \, , 
\label{NnxaU}
\end{equation}
where
\begin{equation}
U(b,b+n+1,x)=x^{-b} \sum_{s=0}^{s=n}  \binom{k}{s} (b)_s x^{-s} \, , 
\label{Kum2}
\end{equation}
is a Kummer polynomial $U (a, b, x)$ in $x$ of degree $n$ when 
$a-b + 1 =-n$, $n \in \mathbb{Z}_{+}$. Generally $U(a,b,x)$
solves Kummer's equation:
\[
x \frac{d^2w}{d^2 x}+(b-x) \frac{d w} {d x} -a w=0 \, .
\]
Starting with $N_0 (x,\mathsf{a})=1$ and applying the first of the basic two
recurrence relations: 
\begin{align}
N_{m+1}(x,\mathsf{a})&= 2x N_m (x,\mathsf{a})+(\mathsf{a}-2) N_m(x,\mathsf{a}-2),
\label{recurNm}\\
2 m N_{m-1}(x,\mathsf{a})&=N_{m}(x,\mathsf{a}+2)-N_m (x,\mathsf{a})
=\frac{d N_{m}(x,\mathsf{a})}{d x}
\quad m=0,1,2,{\ldots} , 
\label{dNmdx}
\end{align}
that can be derived from relation \eqref{NnxaU}, one first obtains  $N_1 (x,a)=2x+a-2$ and eventually arrives at 
a general expression:
\begin{equation}
N_m(x,\mathsf{a})
=\sum_{p=0}^{p=m} \binom{m}{p} (2 x)^{p}
(\mathsf{a}-2)(\mathsf{a}-4)\cdots (\mathsf{a}-2(m-p))\, .
\label{NkDxa}
\end{equation}

For the $T_2^{-1}$ orbit:
\begin{equation}
  \mathbb{T} ( k,0; \mathsf{a})=   T_2^{-k} \vert F=1,\, G=0  
 \rangle_{\alpha_{\mathsf{a}}} \, ,
\label{pureT2k}
\end{equation}
we obtain from expression for the action by $T_2^{-1}$:
\begin{equation}
T_2^{-1} (F,G)= \left(
G-\frac{\gamma_1}{F}+
\frac{1/x-\gamma_2}{1-F+\frac{\gamma_1+\gamma_4}{1-G+\frac{\gamma_1}{F
}}},
1-F+\frac{\gamma_1+\gamma_4}{1-G+\frac{\gamma_1}{F}} \right)\, ,
\label{T2inv}
\end{equation}
a chain of transformations that results 
in a recurrence relation:
\begin{equation}
G^{k}=\frac{2k}{2x(1-G^{k-1})+(\mathsf{a}+2k-2)} \, .
\label{recurG}
\end{equation}
The solution is given this time by
\begin{equation}
G_{k}= 1- \frac{R_{k}(x,\mathsf{a}-2)}{R_{k}(x,\mathsf{a})}= 
\frac{2k R_{k-1}(x,\mathsf{a})}{R_{k}(x,\mathsf{a})}\, ,
\label{GkT2orbit}
\end{equation}
in terms of Kummer polynomials:
\begin{equation}
\begin{split}
&R_n (x,\mathsf{a})=
\sum_{p=0}^{p=n} \binom{n}{p} (2 x)^{p}
\mathsf{a} (\mathsf{a}+2)(\mathsf{a}+4)\cdots (\mathsf{a}a+2(n-p-1))\\
&=2^n x^n x^{\mathsf{a}/2} U(\frac{\mathsf{a}}{2},\frac{\mathsf{a}}{2}+n+1,x) \, ,
\label{Rnxa}
\end{split}
\end{equation}
The above expression can alternatively be obtained 
from the initial condition  $R_0 (x,\mathsf{a})=1$ by 
applying the recursion relation \eqref{Rrecur2} from the two basic
recursive relations satisfied by polynomials $R_k (x,\mathsf{a})$ :
\begin{align}
R_{k+1} (x,\mathsf{a})&= 2 x R_{k} (x,\mathsf{a})+\mathsf{a} R_{k} (x,\mathsf{a}+2)\, ,
\label{Rrecur2}\\
2 k R_{k-1}(x,\mathsf{a})&= R_k(x,\mathsf{a})-R_k(x,\mathsf{a}-2)=
 \frac{d R_k (x,\mathsf{a}) }{d x} \, .
 \label{Rrecur1}
\end{align}

Polynomials denoted by $N_k (x,\mathsf{a})$ for the $T_4$-orbit and $R_k
(x,\mathsf{a})$ for  the $T_2$-orbit, are both of order $k$, and are 
related to each other through relation:
\begin{equation}
R_k (-x,\mathsf{a}) = (-1)^k N_k (x,2-\mathsf{a}) \, .
\label{mxma}
\end{equation}
that is consistent with  $P_1$ reflection that induces $x \to -x , \; 
F \rightarrow 1-G,\; G\rightarrow 1-F$ and $\alpha_1 \leftrightarrow 
\alpha_4,\; \alpha_2 \leftrightarrow 
\alpha_3$. For $\alpha_i$ given in relation \eqref{alphavalues}
this is equivalent to
\begin{equation} {\widehat P}_1 \; :\; k \leftrightarrow m, \; 
\mathsf{a} \leftrightarrow 2-\mathsf{a}, \; x \leftrightarrow -x \, .
\label{sigma4trans}
\end{equation}
\subsection{A  General Case  of $T_4^m T_2^{-k}$ 
orbit and the generalized higher-type Kummer polynomials}
\label{subsection:gencase}
We now consider the full orbit given as in \eqref{orbit} by :
\[ 
T_2^{-k} T_4^{m} \vert F=1,\, G=0  \rangle_{\alpha_{\mathsf{a}}}=
\big( F^{k}_{m} (x,\mathsf{a}), \quad G^{k}_{m} (x,\mathsf{a}) \big), \; \;
\; k, m \in \mathbb{Z}_{+} \,.\]
Here $F^{k}_{m} (x,\mathsf{a}), \, G^{k}_{m} (x,\mathsf{a})$ are solutions to 
equations \eqref{xhameqs} 
\begin{equation}\begin{split}
\frac{d}{d x} F^k_m &=F^k_m (F^k_m -1)(2G^k_m-1)
-\frac{\mathsf{a} + 2 k-2 m}{2x} F^k_m +
\frac{\mathsf{a} + 2 k}{2x} \, ,\\
\frac{d}{d x} G^k_m &= -G^k_m (G^k_m-1)(2F^k_m-1) + 
\frac{\mathsf{a} + 2k-2m-2}{2x} G^k_m
+\frac{-2 k}{2x} \, ,
\label{orbithameqs}
\end{split}
\end{equation}
with parameters given in \eqref{alphavalues}.

See also the reference \cite{ohta}  for use of the 
translation operators to obtain solution of Painlev\'e V equation in
terms of Laguerre polynomials and \cite{CD} for a more recent discussion.

Here we are able to find the Kummer polynomial representation for both
$F^k_m$ and $G^k_m$ as given in the main result of this section by the
following expression: 
\begin{equation}
\begin{split}
F^k_m(x,\mathsf{a})&=
\frac{R_{m k}^{(m-1)} (x,\mathsf{a}-2) 
R_{m(k+1)}^{(m-1)} (x,\mathsf{a})}
{R_{m k}^{(m-1)} (x,\mathsf{a})
R_{m(k+1)}^{(m-1)}  (x,\mathsf{a}-2)}\, ,\\
G^k_m (x,\mathsf{a}) & = 2 k 
\frac{R_{m(k+1)}^{(m-1)} (x,\mathsf{a}-2) R^{(m)}_{(m+1)(k-1)} (x,\mathsf{a}) }
{R_{m k}^{(m-1)} (x,\mathsf{a}-2) R_{(m+1)k}^{(m)}  (x,\mathsf{a})}, \;\; m,k=0,1,2,..
\, .
\label{FGkm}
\end{split}
\end{equation}

The symbols $R_n^{(0)} (x,\mathsf{a})=R_n (x,\mathsf{a})$ are regular Kummer polynomials  of the $n$-th order
defined in relation \eqref{Rnxa}.
The symbols $R_n^{(m)} (x,\mathsf{a})$ are generalized Kummer 
polynomials of the $n$-th order of 
the $m$-th type.
They are obtained from Kummer polynomials by a set of recurrence
relations :
\begin{equation}
\begin{split}
R^{(m-2)}_{(m-1)(k+1)}(x, \mathsf{a}-2) R^{(m)}_{(m+1) k} (x,\mathsf{a})
&=\frac{1}{2m} \big[
R_{m k}^{(m-1)} (x,\mathsf{a}-2) R_{m(k+1)}^{(m-1)} (x,\mathsf{a}) \\&-
R_{m k}^{(m-1)} (x,\mathsf{a}) R_{m(k+1)}^{(m-1)}  (x,\mathsf{a}-2) \big]\, ,
\end{split}
\label{masterrecurrence}
\end{equation}
that
are valid for $m=1,2,3,{\ldots} $ and $k=0,1,{\ldots} $  with initial  
variables being 
$R^{(-1)}_{0}(x, \mathsf{a}-2)=1$ and $ R^{(0)}_{k}(x, \mathsf{a})=R_{k}(x, \mathsf{a})$.
The master recurrence 
relation  \eqref{masterrecurrence} gives rise 
to the 
determinant expressions for the $mk$ order and the $m-1$-type 
Kummer-like polynomials $R^{(m-1)}_{m k} (x,\mathsf{a})$.
Their determinant expressions are obtained by employing 
the Desnanot-Jacobi identity:
\begin{equation}\label{desnajac}
\vert M\vert \times \vert M_{1,n}^{1,n}\vert=\vert M_n^n\vert \times \vert M_1^1\vert -
\vert M_1^n\vert \times \vert M_n^1\vert
\end{equation}
which is valid for any a $(k+1)\times (k+1)$ matrix $M$. 
The notation in identity \eqref{desnajac} is such that
 $M_{i}^{j}$ is a matrix $M$ with
$j$-th row and $i$-th column removed.
When  the Desnanot-Jacobi identity is applied on the above 
recurrence relation \eqref{masterrecurrence}
(up to a constant $C_{m-1}$ determined below) 
one finds $mk$ order and the $m-1$-type 
Kummer-like polynomial $R^{(m-1)}_{m k} (x,\mathsf{a})$ being the determinant  
of the $m \times m$ matrix $M$ with matrix elements to be given by:
\begin{equation}
R_{m k}^{(m-1)} (x, \mathsf{a})= C_{m-1} \vert M \vert \, ,\quad
M_{ij} =R_{k+i-1} (x ,  \mathsf{a}-2(m-j)), \quad i,j=1,{\ldots} ,m, 
\label{gendet}
\end{equation}
where $R_{n} (x ,  b)$ are the original Kummer polynomials \eqref{Rnxa}
and the constants satisfy the relation:
\begin{equation}
C_m =\frac{1}{2m} \frac{C_{m-1}^2}{C_{m-2}}, \quad m=2,3,4,{\ldots} \, .
\label{cmrecursion}
\end{equation}
It follows for the first few cases that
\[
C_0=1, \; C_1= \frac12, \; C_2= \frac{1}{2 \cdot 2} (\frac12)^2=
\frac{1}{16}\, .
\]
Generally, we find from relation \eqref{cmrecursion}:
\begin{equation}
C_m =  \frac{1}{2^{m(m+1)/2}} \prod_{j=0}^{m-2}
\frac{1}{(m-j)^{j+1}}, \quad m=2,3,4,{\ldots} 
\label{cmrecursions}
\end{equation}
The value of $C_m$ is such that the identity $R_0^{(m)}(x,a)=1,
m=0,1,2,{\ldots} $ holds as follows from the above determinant expressions.

From the determinant expression \eqref{gendet} we find as a special 
example $R_{2k}^{(1)}  (x,\mathsf{a}) $  to be given by:
\begin{equation}
R_{2k}^{(1)}  (x,\mathsf{a})= \frac12 \begin{vmatrix} R_{k} (x,\mathsf{a}-2)& 
R_{k} (x,\mathsf{a})\\
R_{k+1} (x,\mathsf{a}-2)&R_{k+1}(x,\mathsf{a})
\end{vmatrix} = \frac12 \bigl( R_{2}(x,\mathsf{a}) R_{1} (x,\mathsf{a}-2)-
R_{1}(x,\mathsf{a}) R_{2} (x,\mathsf{a}-2) \bigr)\,.
\label{R1determinant}
\end{equation}
One can check explicitly that
\begin{equation}
R_{2}^{(1)}  (x,\mathsf{a})  =N_2 (x,\mathsf{a}+2)\,,
\label{R1N2}
\end{equation}
which is a special case of  ``duality''relation \eqref{duality} to be
introduced below and shown to play an important role in  a
general scheme illustrating the power of symmetry generated by
${\widehat P}_1$ automorphism.

Other basic examples of $R_{mk}^{(m-1)}$ are given by 
the following determinant expressions for $m=3$  and $m=4$ :
\begin{align}
R_{3 k}^{(2)}  (x,\mathsf{a})&= \frac{1}{16} 
\begin{vmatrix} R_{k} (x,\mathsf{a}-4)& R_{k} (x,\mathsf{a}-2)& R_{k} (x,\mathsf{a})\\
R_{k+1} (x,\mathsf{a}-4)& R_{k+1} (x,\mathsf{a}-2)& R_{k+1} (x,\mathsf{a})\\
R_{k+2} (x,\mathsf{a}-4)& R_{k+2} (x,\mathsf{a}-2)& R_{k+2} (x,\mathsf{a})
\end{vmatrix}\, \;\; k=1,2,3 ,{\ldots},
\label{R2-3k-determinant}\\
R_{4 k}^{(3)}  (x,\mathsf{a})&= \frac{1}{3} (\frac{1}{16})^2
\begin{vmatrix} R_{k} (x,\mathsf{a}-2 \cdot 3)& R_{k} (x,\mathsf{a}-2 \cdot 2)& R_{k} (x,\mathsf{a}-2)& R_{k} (x,\mathsf{a})\\
R_{k+1} (x,\mathsf{a}-2 \cdot 3)& R_{k+1} (x,\mathsf{a}-2 \cdot 2)& R_{k+1} (x,\mathsf{a}-2)& R_{k+1} (x,\mathsf{a})\\
R_{k+2} (x,\mathsf{a}-2 \cdot 3)& R_{k+2} (x,\mathsf{a}-2 \cdot 2)& R_{k+2} (x,\mathsf{a}-2)& R_{k+2} (x,\mathsf{a})\\
R_{k+3} (x,\mathsf{a}-2 \cdot 3)& R_{k+3} (x,\mathsf{a}-2 \cdot 2)& R_{k+3} (x,\mathsf{a}-2)& R_{k+3} (x,\mathsf{a})
\end{vmatrix}\, .
\label{R3-4k-determinant}
\end{align}
It follows from the above determinant expressions 
that Kummer polynomials 
of the $p$-th kind 
and of the $n$-th order 
are described by a general formula:
\begin{equation}
R_n^{(p)} (x,\mathsf{a})=
\sum_{k=0}^n \binom{n}{k} (2x)^k \left(\mathsf{a}-a_1^{(n,k,p)}\right){\ldots} 
\left(\mathsf{a}-a_{n-k}^{(n,k,p)}\right), \quad p \ge 0, n>0, \;\;R_{0}^{(p)}
(x,\mathsf{a})=1 \, ,
\label{monomials}
\end{equation}
and thus fully defined 
by roots $a_i^{(n,k,p)}$ of the monomials for
$i=1,{\ldots} , n-k$.

Using equation : $d R_n (x,\mathsf{a}) /d x= 
R_n (x,\mathsf{a})-R_n (x,\mathsf{a}-2)$ we find that
\[ \begin{split} 
R_{n} (x,\mathsf{a}-6)&=R_n(x,\mathsf{a})-3 R_n^{\prime} (x,\mathsf{a})+3 R_n^{\prime\prime} (x,\mathsf{a}) -
R_n^{\prime\prime\prime} (x,\mathsf{a}), \\
R_{n} (x,\mathsf{a}-4)&=R_n(x,\mathsf{a})-
2 R_n^{\prime} (x,\mathsf{a})+R_n^{\prime\prime} (x,\mathsf{a}),\;
R_{n} (x,\mathsf{a}-2)=R_n(x,\mathsf{a})-R_n^{\prime} (x,\mathsf{a})\, ,
\end{split} 
\]
and accordingly the determinant in equation \eqref{R3-4k-determinant} can be
rewritten as Wronskian of Kummer polynomials:
\begin{equation}
R_{4 k}^{(3)}  (x,\mathsf{a})= \frac{1}{3} (\frac{1}{16})^2
\begin{vmatrix} R_{k}^{\prime\prime\prime}  (x,\mathsf{a})& R_{k}^{\prime\prime}  (x,\mathsf{a})
& R_{k}^{\prime} (x,\mathsf{a})& R_{k} (x,\mathsf{a})\\
R_{k+1}^{\prime\prime\prime}  (x,\mathsf{a})& R_{k+1}^{\prime\prime}  (x,\mathsf{a})
& R_{k+1}^{\prime} (x,\mathsf{a})& R_{k+1} (x,\mathsf{a})\\
R_{k+2}^{\prime\prime\prime}  (x,\mathsf{a})& R_{k+2}^{\prime\prime}  (x,\mathsf{a})
& R_{k+2}^{\prime} (x,\mathsf{a})& R_{k+2} (x,\mathsf{a})\\
R_{k+3}^{\prime\prime\prime}  (x,\mathsf{a})& R_{k+3}^{\prime\prime}  (x,\mathsf{a})
& R_{k+3}^{\prime} (x,\mathsf{a})& R_{k+3} (x,\mathsf{a})
\end{vmatrix} \, ,
\label{R3-4k-wronskian}
\end{equation}
with all polynomials now taken at the same point $\mathsf{a}$.

There also exists a dual master recurrence relation:
\begin{equation}
\begin{split}
&\frac{1}{2k} \left( 
R_{m k}^{(m-1)} (x,\mathsf{a}-2) R_{(m+1)k}^{(m)}  (x,\mathsf{a})
-R_{m k}^{(m-1)} (x,\mathsf{a}) 
R_{k(m+1)}^{(m)} (x,\mathsf{a}-2)\right)\\&=
R_{m(k+1)}^{(m-1)} (x,\mathsf{a}-2) R^{(m)}_{(k-1)(m+1)} (x,\mathsf{a}) \, .
\label{dualmaster}
\end{split}
\end{equation}
Applying the duality relation
\begin{equation}
R_{k m }^{(m-1)} (x,\mathsf{a}-2) {=}  N_{k m}^{(k-1)} (x,\mathsf{a})\, ,
\label{duality}
\end{equation}
that exchanges $m \leftrightarrow k$ and $\mathsf{a} \to \mathsf{a}-2$ and introduces
the higher type order generalization of Kummer polynomials 
$ N_{k } (x,\mathsf{a})$ defined in \eqref{NnxaU}
we obtain the dual version of the recurrence relations
\eqref{masterrecurrence} :
\begin{equation}
\begin{split}
&\frac{1}{2k} \left( N_{m k}^{(k-1)} (x,\mathsf{a}) N_{(m+1)k}^{(k-1)}  (x,\mathsf{a}+2)
-N_{m k}^{(k-1)} (x,\mathsf{a}+2) 
N_{k(m+1)}^{(k-1)} (x,\mathsf{a})\right)\\&=
N_{m(k+1)}^{(k)} (x,\mathsf{a}) N^{(k-2)}_{(k-1)(m+1)} (x,\mathsf{a}+2)\, ,
\label{dualmastera}
\end{split}
\end{equation}
that via  Desnanot-Jacobi identity \eqref{desnajac}
leads to expression for the generalized higher type Kummer polynomials 
$ N_{k } (x,\mathsf{a})$ :
\begin{equation}
N_{km}^{(k-1)}(x,\mathsf{a})=C_{k-1} \vert M\vert, \quad M_{ij}=N_{m+i-1} (x,
\mathsf{a}+2(j-1))\, ,
\label{Ndeterminant}
\end{equation}
where the normalization constant $C_{k-1}$ is defined in \eqref{cmrecursion}.
For example
\begin{equation}
N_{3 m}^{(2)}  (x,(x,\mathsf{a}))= \frac{1}{16} 
\begin{vmatrix} N_{m} (x,\mathsf{a})& N_{m} (x,\mathsf{a}+2)& N_{m} (x,\mathsf{a}+4)\\
N_{m+1} (x,\mathsf{a})& N_{m+1} (x,\mathsf{a}+2)& N_{m+1} (x,\mathsf{a}+4)\\
N_{m+2} (x,\mathsf{a})& N_{m+2} (x,\mathsf{a}+2)& N_{m+2} (x,\mathsf{a}+4)
\end{vmatrix}\, \;\; m=1,2,3 ,{\ldots} 
\label{N2-3k-determinanta}
\end{equation}

Thus we have determinant expressions \eqref{gendet}
and \eqref{Ndeterminant} that are dual to each other
according to the duality relation \eqref{duality} that can also be viewed
as generated by ${\widehat P}_1$ transformation
$m \leftrightarrow k$, $ x \to -x,  \mathsf{a}\to 2-\mathsf{a}$.
In view of the duality relations \eqref{duality} we have 
the following relation
\begin{equation}
R_{k m}^{(m-1)} (-x,-\mathsf{a})= (-1)^{km}  R_{k m}^{(k-1)} (x,\mathsf{a}). \,
\label{dualityRRxa}
\end{equation}
To understand better relation \eqref{dualityRRxa} it  is illustrative to introduce 
notation $R_{m,k} (x,\mathsf{a})=R_{k m}^{(m-1)} (x,\mathsf{a})$. In this new notation 
the relation \eqref{dualityRRxa} describes a condition for commutation of indices
$m,k\to k, m$ : $R_{k,m}  (x,\mathsf{a})=  (-1)^{km}R_{m,k}  (-x,-\mathsf{a})$.

Also, thanks to duality relation \eqref{duality}, we have an alternative
 expression for our solutions as :
\begin{equation}
\begin{split}
F^k_m (x,\mathsf{a})&=
\frac{N_{m k}^{(k-1)} (x,\mathsf{a}) 
N_{m(k+1)}^{(k)} (x,\mathsf{a}+2)}
{N_{m k}^{(k-1)} (x,\mathsf{a}+2)
N_{m(k+1)}^{(k)}  (x,\mathsf{a})}\, ,
\\
G^k_m(x,\mathsf{a}) & =  
 2 k 
\frac{N_{m(k+1)}^{(k)} (x,\mathsf{a}) N^{(k-2)}_{(k-1)(m+1)} (x,\mathsf{a}+2) }
{N_{m k}^{(k-1)} (x,\mathsf{a}) N_{(m+1)k}^{(k-1)}  (x,\mathsf{a}+2)}\,.
\label{FGkma}
\end{split}
\end{equation}

\section{Even ${\widehat P}_2,{\widehat P}_4$ reflections  and degeneracy}
\label{section:dege}
General remark that applies to automorphisms ${\widehat P}_4,{\widehat
P}_2, {\widehat P}_3$ is  that 
implementing them requires enforcing the constant $a/2$ to be an
integer, chosen here to be  $l \in \mathbb{Z}$.

\subsection{Action of the ${\widehat P}_2$ automorphism}
\label{subsection:sigma2}
For ${\widehat P}_2$ the transformation
$\alpha_2 \leftrightarrow \alpha_4 $ for $\alpha_2=-2k$ and 
$\alpha_4=2-\mathsf{a} +2m =2(1-l+m)$ is equivalent to 
that of
\[ %
l \to  1+m+k, \;\;\; k \to l-1-m, %
\]
for $l>m$.

Accordingly  ${\widehat P}_2$ transformations are such 
that \[\begin{split}
{\widehat P}_2 (F_m^k (x,l))&= %
F_m^{l-1-m}  (-x, 1+m+k)\\
{\widehat P}_2 (G_m^k (x,l))&= %
G_m^{l-1-m}  (-x, 1+m+k)
, \quad \mathsf{a}=2l
\end{split}
\]
and are expected to be solutions to equations
\eqref{xhameqs} with the same parameters as
solutions $F_m^k (x,\mathsf{a}), 1- G_m^k (x,\mathsf{a})$. Are these functions equal to
the original solutions $F_m^k (x,\mathsf{a}), 1- G_m^k (x,\mathsf{a})$ or are they
degenerated solutions is 
the main question. We will fully
answer it in the subsection \ref {subsection:degerefl} but first we
will discuss how it works on an explicit example:
\begin{exmp}
Here we calculate a solution $T_2^{-k} T_4^m \vert F=1,\, G=0  
\rangle_{\alpha_{\mathsf{a}=4}} $
using the expression for solutions
found in equation \eqref{FGkm} for $k=m=1$ with $l=2$ ( $\mathsf{a}=2 l=4$): 
\begin{equation}
\begin{split}
F_{1}^{1} (x,\mathsf{a}=4)&=
\frac{R_{1} (x,\mathsf{a}-2) 
R_{2} (x,\mathsf{a})}
{R_{1} (x,\mathsf{a})
R_{2}  (x,\mathsf{a}-2)} \vert_{\mathsf{a}=4} = 
\frac{(2x+2)((2x)^2+16x+24)}
{(2x+4)((2x)^2+8x+8)}
\\
G_{1}^{1}(x,\mathsf{a}=4) & = 2  
\frac{R_{2} (x,\mathsf{a}-2) R^{(1)}_{0} (x,\mathsf{a}) }
{R_{1} (x,\mathsf{a}-2) R_{2}^{(1)}  (x,\mathsf{a})}\vert_{\mathsf{a}=4} =
2 \frac{R_{2} (x,\mathsf{a}-2) }
{R_{1} (x,\mathsf{a}-2) N_{2}  (x,\mathsf{a}+2)}\vert_{\mathsf{a}=4}\\
&= 2  
\frac{((2x)^2+8x+8)}
{(2x+2)((2x)^2+16x+8)} \, ,
\label{FGk1m1}
\end{split}
\end{equation}
where we used the identities \eqref{R1N2} and $R_{0}^{(m)}  (x,\mathsf{a})=1$.
The above two solutions %
in relations \eqref{FGk1m1}  satisfy the equations \eqref{orbithameqs}
with $k=m=1$, ($\alpha_1=\mathsf{a}+2=6,\alpha_2=-2,\alpha_3=-2$). We write these 
parameters as: 
\begin{equation}
\alpha_i= 2 (k+l,-k,-m, 1-l+m)=2 (3,-1,-1,0), %
\label{alphakml}
\end{equation}
Under the ${\widehat P}_2$ transformation the parameter $m$ is invariant 
and we have the following expressions for the $\mathsf{a}$, $k$ transformations :
\[ \begin{split}  \bar{ \mathsf{a}}&= \mathsf{a}+2(1+k+m-l)= 4+2=6\\
{\bar k}&=k-(1+k+m-l) =1-1=0 
\end{split} 
\]
Thus we are looking for the transformed solutions given by:
\[\begin{split} 
F^{{\bar k}}_m (-x, \bar{ \mathsf{a}}/2)&= F^0_1 (-x, 6), \\
 G^{{\bar k}}_m (-x, \bar{ \mathsf{a}}/2)&=G^0_1 (-x, 6)\,,
\end{split} 
\]
which we calculate  %
using equation \eqref{FGkm} to obtain:
\begin{equation}
\begin{split}
F^{0}_{1} (-x,6)&=
\frac{R_{0} (-x,4) 
R_{1} (-x,6)}
{R_{0} (-x,6)
R_{1}  (-x,4)}=\frac{6-2x}{4-2x}= \frac{x-3}{x-2}\, ,\\
G^{0}_{1}(-x,6) & = 2 \cdot 0 \cdot  
\frac{R_{1} (-x,4) R_{-2)}^{(1)} (-x,6) }
{R_{0} (-x,4) R_{0}^{(1)}  (-x,6)} =0\, .
\label{FGkm10}
\end{split}
\end{equation}

One checks explicitly that $F=(x-3)/(x-2)$ and
$G=1$ solve  the equation \eqref{xhameqs} with
$\alpha_1=6, \alpha_2=-2, \alpha_3=-2$ that agrees with
the parameters \eqref{alphakml} of the original solution.
We will see in subsection \ref{subsection:degerefl} that this degeneracy
is a general feature of the even ${\widehat P}_{2i}$ automorphisms.

\end{exmp}

Generally we can rewrite
\begin{equation}
\begin{split}
F_m^{l-1-m}  (-x, 2(1+m+k))&=
\frac{R_{m (l-1-m)}^{(m-1)} (-x,2(k+m)) 
R_{m(l-m)}^{(m-1)} (-x,2(1+m+k))}
{R_{m (l-1-m)}^{(m-1)} (-x,2(1+m+k))
R_{m(l-m)}^{(m-1)}  (-x,2(k+m))}\\
&=
\frac{N_{m (l-1-m)}^{(m-1)} (x,2(1-k-m)) 
N_{m(l-m)}^{(m-1)} (x,-2(m+k))}
{N_{m (l-1-m)}^{(m-1)} (x,-2(m+k))
N_{m(l-m)}^{(m-1)}  (x,2(1-k-m))}
\end{split}
\label{degeF}
\end{equation}
where we used the identity
\begin{equation}
R_{m k}^{(m-1)} (x,\mathsf{a})= (-1)^{m k}N_{m k}^{(m-1)} (-x,2-a), \;\;
R_{m k}^{(m-1)} (-x,a)= (-1)^{m k}N_{m k}^{(m-1)} (x,2-a), \;\;
\label{umidentity}
\end{equation}
Expression \eqref{degeF} describes a closed expression for the
degenerated partner of :
\[
F^k_m (x,2l)=
\frac{R_{m k}^{(m-1)} (x,2(l-1) 
R_{m(k+1)}^{(m-1)} (x,2l)}
{R_{m k}^{(m-1)} (x,2l)
R_{m(k+1)}^{(m-1)}  (x,2(l-1)}\, ,
\]
which is a solution  given in \eqref{FGkm} for $m,k, a =2l$.

For the the remaining part of solution we find that 
the degenerated solution is:
\begin{equation}
\begin{split}
&G_m^{l-1-m}  (-x, 2(1+m+k))\\&=
2 (l-1-m)
\frac{R_{m(l-m)}^{(m-1)} (-x,2(m+k)) R^{(m)}_{(l-m-2)(m+1)} (-x, 2(1+m+k) ) }
{R_{m (l-1-m)}^{(m-1)} (-x,2(m+k)) R_{(m+1)(l-1-m)}^{(m)}
(-x,2(1+m+k))} \\
&=
-2 (l-1-m)\frac{N_{m(l-m)}^{(m-1)} (x,2(1-m-k)) 
N^{(m)}_{(l-m-2)(m+1)} (x, -2(m+k) ) }
{N_{m (l-1-m)}^{(m-1)} (x,2(1-m-k)) N_{(m+1)(l-1-m)}^{(m)}
(x,-2(m+k))} \, ,
\label{degeG}
\end{split}
\end{equation}
where we again used  equation \eqref{umidentity}. 

Remember that 
${\widehat P}_2(G)=1-G_m^{l-1-m}  (-x, 2(1+m+k))$. Thus
\begin{equation}
{\widehat P}_2(G) = 1+2 (l-1-m)\frac{N_{m(l-m)}^{(m-1)} (x,2(1-m-k)) 
N^{(m)}_{(l-m-2)(m+1)} (x, -2(m+k) ) }
{N_{m (l-1-m)}^{(m-1)} (x,2(1-m-k)) N_{(m+1)(l-1-m)}^{(m)}
(x,-2(m+k))} \,.
\label{degeGa}
\end{equation}
Together with \eqref{degeF}, formula \eqref{degeGa} solves \eqref{xhameqs}
with the identical parameters
$\alpha_i= 2 (l+k,-k,-m,1-l+m)$ as the original solution \eqref{FGkm}
with $\mathsf{a}=2l$.

\begin{exmp}
\label{exmp:n2=2n4=2}
We consider the case of  
\begin{equation}   k=1 ,\,  m=1,\, l=4,\, a=2\cdot l=8
\,\to \, \alpha_i=2 (l+k,-k,-m,1-l+m)=2 (5,-1,-1,-2)\, ,
\label{itemA}
\end{equation}
We find ${\bar k}= {l-1-m}=2$, ${\bar l}=(1+m+k)=3$.
Applying the formulas \eqref{degeF} and \eqref{degeGa} 
we obtain the ${\widehat P}_2$ transformed degenerated solutions :
\[
\begin{split}
F^{2}_1 (-x, 2 \cdot 3)&=
\frac{N_{2} (x,- 2 \cdot 1) 
N_3 (x,-2\cdot 2)}
{N_{2} (x,-2\cdot 2)
N_{3}  (x,- 2\cdot 1)}\\
&= \frac{(x^2-4 x+6) (x^3-9x^2+36 x-60)}{(x^2-6 x+12) (x^3-6 x^2+18
x-24)}\\
1-G^{2}_1 (-x, 2 \cdot 3)&=
1-(-2 \cdot 2)\frac{N_{3} (x,-2)) 
N^{(1)}_{2} (x, -4 ) }
{N_{2} (x,-2) N_{4}^{(1)}
(x,-4)} \\&=
\frac{ (x^2-6 x+12) (x^4-8 x^3+24 x^2-24 x+12)}
{(x^2-4 x+6) (72+54 x^2+x^4-12 x^3-96 x)} \, .
\end{split}
\]
Inserting $x=-z^2/2$ and multiplying by $z$ we obtain from the above
two results the expressions (3.15)
for $q(z),p(z) $ of Example 3.1 of \cite{AGLZ2024}
that also solve the relevant Painlev\'e  with the parameters
$\alpha_i=2 (5,-1,-1,-2)$.
\end{exmp}

\subsection{Action of the ${\widehat P}_4$ automorphism}
\label{subsection:sigma1}
The ${\widehat P}_4$ generated transformation
$\alpha_1 \leftrightarrow \alpha_3 $ for $\alpha_1=2(l+k)$ and 
$\alpha_3=-2m$ is equivalent to 
that of
\[ l \to -k-m, \; k \to k, \quad m \to  -l-k\,.
\]
Thus it must hold $-l-k \ge 0$ or $l \le -k$.

Accordingly, 
we obtain:
\begin{equation}\begin{split}
{\widehat P}_4( F_m^k(x,2 l))&=  F_{-l-k}^k(-x,2(-k-m)),\;\\ %
{\widehat P}_4( G_m^k(x,2l))&= G_{-l-k}^k(-x,2(-k-m)) %
\end{split}\label{P4FG}
\end{equation}
The first of equations \eqref{P4FG} yields after
substituting the relation \eqref{FGkm}
for $G_{m}^{k} (x,\mathsf{a})$:
\[
\begin{split}G_{-l-k}^{k} (-x,-k-m) &= 2 k \frac{R_{(-l-k)(k+1)}^{(-l-k-1)} 
(-x,2(-k-m-1))
R_{(-l-k+1)(k-1)}^{(-l-k)} (-x,2(-k-m))}
{R_{(-l-k) k}^{(-l-k-1)} (-x,2(-k-m-1)) R_{(-l-k+1)k}^{(-l-k)}
(-x,2(-k-m))} \\
&=-2 k \frac{N_{(-l-k)(k+1)}^{(-l-k-1)} (x,2(2+k+m))
N_{(-l-k+1)(k-1)}^{(-l-k)} (x,2(1+k+m))}
{N_{(-l-k) k}^{(-l-k-1)} (x,2(2+k+m)) N_{(-l-k+1)k}^{(-l-k)}
(x,2(1+k+m))}
\end{split}
\]
From ${\widehat P}_4(F)=F^{k}_{-l-k} (-x, -2(k+m))$ %
we get
\[ \begin{split}
F^{k}_{-l-k} (-x, -2(k+m))&=\frac{R_{(-(l-k) k}^{(-l-k-1)} (-x,-2(1+k+m)) 
R_{(-l-k)(k+1)}^{(-l-k-1)} (-x,-2(k+m))}
{R_{(-l-k) k}^{(-l-k-1)} (-x,-2(k+m))
R_{(l-k)(k+1)}^{(-l-k-1)}  (-x,-2(1+k+m))}\\
&= \frac{N_{(-(l-k) k}^{(-l-k-1)} (x,2(2+k+m)) 
N_{(-l-k)(k+1)}^{(-l-k-1)} (x,2(1+k+m))}
{N_{(-l-k) k}^{(-l-k-1)} (x,2(1+k+m))
N_{(l-k)(k+1)}^{(-l-k-1)}  (x,2(2+k+m))}
\end{split}
\]
We will apply these results in subsection \ref{subsection:degerefl}
and Example \ref{example:p4},
where we will study degeneracy introduced by even reflection automorphisms
and answer the question whether the results for ${\widehat P}_4(F_m^k(x,2l))$
and ${\widehat P}_4( G_m^k(x,2l))$ are equal to
$1-F_m^k(x,2l)$ and $G_m^k(x,2l)$ that share the same parameters.

\subsection{Degeneracy induced by even reflections 
${\widehat P}_2$ and ${\widehat P}_4$}
\label{subsection:degerefl}
We now present a general discussion 
on ${\widehat P}_4,{\widehat P}_2$ reflections that
acting on solutions $T_2^{-k}T_4^m  \vert g_1=1,g_2=0\rangle_a$
produce degenerated solutions (solutions that share 
the same parameters $\alpha_i$ as original solutions).

The result will follow if we are able to show the following equalities:
\begin{align}
{\widehat P}_2 T_2^{-k_3}T_4^{m_3}  \vert
g_1=1,g_2=0\rangle_{a_3}
&{=}\pi s_1 T_2^{-k_1}T_4^{m_1}  \vert
g_1=1,g_2=0\rangle_{a_1} \, ,\label{R2pis}\\
{\widehat P}_4 T_2^{-k_4}T_4^{m_4}  \vert
g_1=1,g_2=0\rangle_{a_4}
&{=}\pi^{-1} s_4 T_2^{-k_2}T_4^{m_2}  \vert
g_1=1,g_2=0\rangle_{a_2} \, ,	
\label{R0pis}
\end{align}
for some values of $k_i, m_i$ and $a_i$, since  we have previously
established in \cite{AGLZ2024} that both solutions on the right hand sides are different from
$T_2^{-k}T_4^{m} \vert g_1=1,g_2=0\rangle_a$, while they %
share the identical parameters.

First,  we will show that the two relations given in equations
\eqref{R2pis} and \eqref{R0pis}
are equivalent. Assume that equation \eqref{R2pis} holds and
multiply the equation \eqref{R2pis} by ${\widehat P}_2$. 
Due to  relation $\pi = {\widehat P}_2 {\widehat P}_1$
we  obtain that equation \eqref{R2pis} is equivalent to:
\[
\begin{split}
T_2^{-k_3}T_4^{m_3}  \vert
g_1=1,g_2=0\rangle_{a_3}
&={\widehat P}_1 s_1 T_2^{-k_1}T_4^{m_1}  \vert
g_1=1,g_2=0\rangle_{a_1}\\
&=s_4 {\widehat P}_1 T_2^{-k_1}T_4^{m_1}  \vert
g_1=1,g_2=0\rangle_{a_1}
\end{split}
\]
where in the last equation we used that ${\widehat P}_1 s_1 {\widehat P}_1=s_4$.
Multiplying both sides by $s_4$ and then using that
${\widehat P}_1{\widehat P}_2=\pi^{-1}$ we get :
\[
\begin{split}
\pi^{-1} s_4 {\widehat P}_2 T_2^{-k_3}T_4^{m_3}  \vert
g_1=1,g_2=0\rangle_{a_3}
&= {\widehat P}_1{\widehat P}_2 {\widehat P}_1 s_1 T_2^{-k_1}T_4^{m_1}  \vert
g_1=1,g_2=0\rangle_{a_1}\\
&={\widehat P}_4 T_2^{-k_1}T_4^{m_1}  \vert
g_1=1,g_2=0\rangle_{a_1}\, ,
\end{split}
\]
where we used the identity ${\widehat P}_1{\widehat P}_2 {\widehat P}_1={\widehat P}_4$ that follows from
\eqref{conjugation}. Thus  
equation \eqref{R0pis} follows from  equation  \eqref{R2pis}  and we
will only need to show that one of these two equations holds.

We now embark on a proof of equation  \eqref{R2pis}.
First,  we  recall that
\[\begin{split}{\widehat P}_2 \vert g_1=1,g_3=0,g_2=0,g_4=1\rangle
&=
\vert g_1=1,g_3=0,g_4=0,g_2=1\rangle, \\
{\widehat P}_2(a,0,0,2-a)&=(a,2-a,0,0)\, ,
\end{split}
\]
then using \eqref{R2translation} we get 
\begin{equation}\begin{split}
&{\widehat P}_2 T_2^{-k_3}T_4^{m_3}  \vert
g_1=1,g_3=0,g_2=0,g_4=1\rangle_{a_3}\\
&= T_1^{k_3}T_3^{-m_3} \vert
g_1=1,g_3=0,g_4=0,g_2=1\rangle_{(a_3,2-a_3,0,0)}\,.
\label{R2TT}\end{split}
\end{equation}
Next we consider parameters of solution \eqref{R2TT} 
 with $ k_3, m_3$ and $a_3$:
\[
T_1^{k_3}T_3^{-m_3} (a_3,2-a_3,0,0) =(a_3+2
k_3,2-a_3+2m_3,-2m_3,-2k_3) \, ,
\]
that will agree with the parameters $(a+2k,-2k, -2m, 2-a+2m)$
of the solution $T_2^{-k}T_4^m  \vert g_1=1,g_3=0,g_2=0,g_4=1\rangle$
if $a=2l$, $a_3= 2 l_3$ and if it holds that
\begin{equation}
k_3= l-m-1, \quad l_3=1+m+k,\quad m_3=m \, .
\label{klm3}
\end{equation}
We will see that for these values we have a degeneracy.

Let us  compare the  two sides of expressions \eqref{R2pis}.
Since we have an identity $\pi={\widehat P}_2 {\widehat P}_1$ we can rewrite  the
equation \eqref{R2pis} as
\[ {\widehat P}_1 s_1 T_2^{-k_1}T_4^{m_1}  \vert g_1=1,g_2=0\rangle_{a_1}
= T_2^{-k_3}T_4^{m_3}  \vert g_1=1,g_2=0\rangle_{a_3}\, ,
\]
or 
\begin{equation}
s_1 T_2^{-k_1}T_4^{m_1}  \vert g_1=1,g_2=0\rangle_{a_1}=
 {\widehat P}_1T_2^{-k_3}T_4^{m_3}  \vert g_1=1,g_2=0\rangle_{a_3}
 = T_4^{k_3}T_2^{-m_3}{\widehat P}_1\vert g_1=1,g_2=0\rangle_{a_3}\,.
 \label{s1t24}
 \end{equation}
The left hand side  of equation \eqref{s1t24} can be rewritten as
\begin{equation}\begin{split}
&T_1^{-k_1}T_4^{m_1} s_1 \vert g_1=1,g_2=0\rangle_{a_1}
=T_1^{-k_1}T_4^{m_1} s_1 T_1^{a_1/2}\vert g_1=1,g_2=0\rangle_{a=0}\\
&=T_1^{-k_1}T_4^{m_1}T_2^{a_1/2} s_1\vert g_1=1,g_2=0\rangle_{a=0}
=T_1^{-k_1}T_4^{m_1}T_2^{a_1/2} \vert g_1=1,g_2=0\rangle_{(0,0,0,2)}\\
&= T_1^{-k_1-1}T_4^{m_1}T_2^{a_1/2} T_1 \vert
g_1=1,g_2=0\rangle_{(0,0,0,2)}\, ,
\end{split}
\label{lefths1}
\end{equation}
where we used the relations $s_iT_i s_i=T_{i+1}$ , $s_i
T_j=T_js_i, j \ne i, i+1$ that hold between the $s_i$
transformations and the translation operators \cite{AGLZ2023}. 
Also we used that $s_1$ is an identity when acting on 
$\vert g_1=1,g_2=0\rangle_{(0,0,0,2)}$. 

On the right hand side of equation \eqref{s1t24} we have
\begin{equation} T_4^{k_3}T_2^{-m_3}{\widehat P}_1\vert g_1=1,g_2=0\rangle_{a_3}
= T_4^{k_3}T_2^{-m_3}T_1^{-a_3/2} \vert g_1=0,g_2=1\rangle_{(2,0,0,0)}
\, ,
\label{righths1}
\end{equation}
where we used that ${\widehat P}_1 T_1 {\widehat P}_1= T_1^{-1}$ and
\[\begin{split}
{\widehat P}_1 (g_1=1,g_2=0,g_3=0,g_4=1)&=(g_4=1,g_3=0,g_2=0,g_1=1)=(1,0,0,1), \\
{\widehat P}_1 (a,0,0,2-a)&= (2-a,0,0,a)\, .
\end{split}
\]
Comparing the powers of $T_1,T_4,T_2$ in expressions \eqref{lefths1}
and \eqref{righths1} we obtain three conditions:
\begin{equation}
l_3=1+k_1, \;\; m_1=k_3, \;\; l_1 =m_3\, ,
\label{3vinculos}
\end{equation}
with $a_i=2l_i, i=1,3$.  For these conditions being satisfied the relation
\eqref{R2pis} holds both for the parameters and the solutions provided that 
\[T_1 \vert
g_1=1,g_2=0\rangle_{(0,0,0,2)}={\widehat P}_1 \vert
g_1=1,g_2=0\rangle_{(0,0,0,2)}=\vert g_1=1,g_2=0\rangle_{(2,0,0,0)}\, ,
\]
but since $T_1$ only changes the parameters
 but not the underlying 
$F  , G$ variables, this relation is an identity  and therefore also the relation
\eqref{R2pis} holds.
Comparing relations \eqref{klm3} and \eqref{3vinculos}
we find a direct relation:
\[ l-m-1=m_1, \; k_1=m+k, \; m=l_1\, ,\]
 between the parameters  $k_1,m_1,l_1$ and 
$k,m,l$.
These are the exact values \cite{AGLZ2024} for which it holds 
that solutions $\pi s_1  T_2^{-k_1} T_4^{m_1} \vert F=1,\, G=0  
\rangle_{\alpha_{\mathsf{b}}}$ and $T_2^{-k} T_4^{m} \vert F=1,\, G=0  
\rangle_{\alpha_{\mathsf{a}}}$ are different while the
four parameters of the two solutions are equal \cite{AGLZ2024}:
\[
\pi s_1  \left(\alpha_{k_1,m_1; \mathsf{b}}\right) = \alpha_{k,m ;
\mathsf{a}}\, .
\]
\begin{exmp}
\[
{\widehat P}_2: (F,G) \to (\text{degenerate} (F), 
\text{degenerate} (G))
\]

Let $m=2$ and $k=1$ and $l=3$, the corresponding solution is
obtained from relation \eqref{FGkm} to be 
\[
\begin{split}
&T_2^{-1} T_4^2  \vert F=1,\, G=0  
\rangle_{\alpha_{\mathsf{a}=6}}=
\bigg(F =
\frac{(x^2+4 x+6) (x^3+9 x^2+36 x+60)}{(x^2+6 x+12) (18 x+6 x^2+x^3+24)}\\&,
G = 2 \frac{(x^2+6x+6) (18 x+6 x^2+x^3+24)}{(x^2+4 x+6) (12 x^3+54
x^2+96 x+72+x^4)} \bigg)\,,
\end{split}
\]
with $\alpha_i= 2 (l+k,-k,-m,1-l+m)=2(4,-1,-2,
0)=2\pi s_1(l_1+k_1,-k_1,-m_1,1-l_1+m_1)$.
The unique solution for $l_1,k_1,m_1$ is:  $l_1=-2, k_1=3, m_1=0$. 
We use  the right hand side of relation \eqref{R2pis}
to write the  degenerated solution as :
\begin{equation}\begin{split}
&\pi s_1 T_2^{-3}  \vert F=1,\, G=0  
\rangle_{\alpha_{\mathsf{a}=-4}}=
\big( \frac{-8x+12+ x^2}{-6x+6+ x^2},\; 1\,\big)\,, \\
&\pi s_1 2 (l_1+k_1,-k_1,-m_1,1-l_1+m_1)=2 \pi s_1
(1,-3,0,3)=2(4,-1,-2,0)\, .
\end{split}
\label{deg-m2k1l3}
\end{equation}

We will now  reproduce the degenerated solution \eqref{deg-m2k1l3} using the
${\widehat P}_2$ transposition and the left hand side of relation
\eqref{R2pis}.
Starting again as  above with  $m=2,k=1$ and $a=6$ or $l=3$ so that $\alpha_i= 2 (l+k,-k,-m,1-l+m)=2(4,-1,-2, 0)$
with  $l-1-m=0$ and $F_m^{l-1-m}  (-x, 2(1+m+k))=F_2^0 (-x,2 \cdot 4)$:
\[\begin{split}
F_2^0 (-x,2 \cdot 4)
&=
\frac{N_{0}^{(1)} (x,2(-2)) 
N_{2(3-2)}^{(1)} (x,-2(3))}
{N_{0}^{(1)} (x,-2(3))
N_{2(3-2)}^{(1)}  (x,2(-2))}=\frac{
N_{2}^{(1)} (x,-6)}
{N_{2}^{(1)}  (x,-4)}\\&=
\frac{-32x+48+4 x^2}{-24x+24+4 x^2}=\frac{-8x+12+ x^2}{-6x+6+ x^2} \, .
\end{split}
\]
Since $l-1-m=0$ then $G_m^{l-1-m}  (-x, 2(1+m+k))=0$, but then 
the ${\widehat P}_2$ transformed solution is ${\widehat P}_2(G)=1
-G_m^{l-1-m}  (-x, 2(1+m+k))$ which is equal to $1$.

The above functions $F_2^0 (-x,2 \cdot 4)=(-8x+12+ x^2)/(-6x+6+ x^2)$
and ${\widehat P}_2(G)=1$ solve the Hamiltonian equations \eqref{xhameqs}
with $\alpha_1=8,\alpha_2=-2, \alpha_3=-4$ coefficients
so that again we obtain acting with  ${\widehat P}_2$ automorphism
a solution for the same parameters 
$\alpha_i=2(4,-1,-2, 0)$ as original $F^k_m, G^k_m$ solution.
\end{exmp}

\begin{exmp}
\label{example:p4}
Here we illustrate how the action of ${\widehat P}_4$ generates
degenerated solutions according to a simple prescription:
\[
{\widehat P}_4: (F,G) \to (\text{degenerate} (F), 
\text{degenerate} (G))\, .
\]
Recall that as shown in subsection \eqref{subsection:sigma1}
we have
\[
{\widehat P}_4: \alpha_1 \leftrightarrow \alpha_3 \; \sim \; 
 l \to %
-m-k\in \mathbb{Z}_{+}, \; m \to %
-l-k\in \mathbb{Z}_{+}, %
\]
and 
\[\begin{split}
{\widehat P}_4 (F_m^k (x,l))&= %
1-F_{-l-k}^{k}  (-x, -m-k)\, ,\\
{\widehat P}_4 (G_m^k (x,l))&= %
G_{-l-k}^{k}  (-x, -m-k)\, 
, \quad \mathsf{a}=2l\,.
\end{split}
\]
Accordingly, for $m=1,k=1, l=-1$ or $\mathsf{a} =-2$ we obtain 
\[\begin{split}
F^1_1 (x,-2)=\frac{(x-2)^2 x}{(x-1)(x^2-4 x+2)}& \;
\stackrel{{\widehat P}_4}{\longrightarrow}\; 1-F_{0}^{1}  (-x, -2)=0\, ,\\
G^1_1(x,-4) = \frac{(x^2-4 x+2)}{(x-2) (x^2-2 x+2)}&\;
\stackrel{{\widehat P}_4}{\longrightarrow}\; G_{0}^{1}  (-x, -2)=
-\frac{1}{x+2}\,.
\end{split}
\]
Both above results are solutions of the $F,G$ equations \eqref{xhameqs} with the same parameters.
\end{exmp}

Although equations \eqref{R2pis} and \eqref{R0pis} follow from each
other they hold for different values of the parameter $l$ such that 
$\mathsf{a} =2l$. To show this we discuss the equality between parameters shared
between the left hand side of equation \eqref{R2pis} and the parameter
of the solution  $T_2^{k}T_4^m \vert g_1=1,g_2=0\rangle_a$. The common
parameters are :
\[T_2^{-k}T_4^m (\mathsf{a},0,0,2-\mathsf{a})={\widehat P}_2 T_2^{-k_2}T_4^{m_2}
(\mathsf{a}_2,0,0,2-\mathsf{a}_2)\,,
\]
or 
\[\begin{split} (l+k,-k,-m, 1-l+m)&={\widehat P}_2(l_2+k_2,-k_2,-m_2,1-l_2+m_2)\\
&=(l_2+k_2,1-l_2+m_2,-m_2,-k_2)\,,
\end{split}
\]
the solution to these conditions is 
\[  m_2=m , l= k_1+m+1,  l_2= k+m_2+1=k+m+1\,,
\]
which implies  that both $l$ and $l_2$ are positive.
\[l>m, l_2>m\, .
\]
Repeating the same analysis for the degeneracy connected with ${\widehat P}_4$ we
start with equality
\[T_2^{-k}T_4^m (\mathsf{a},0,0,2-\mathsf{a})={\widehat P}_4 T_2^{-k_0}T_4^{m_0}
(\mathsf{a}_0,0,0,2-\mathsf{a}_0)\,,
\]
that can be rewritten as 
\[\begin{split} (l+k,-k,-m, 1-l+m)&={\widehat
P}_4(l_0+k_0,-k_0,-m_0,1-l_0+m_0)\\
&=(-m_0, -k_0, l_0+k_0,1-l_0+m_0)\,.
\end{split}
\]
This time it is the $k$ parameter which remains unchanged 
$k=k_0$ while the expressions for $l,l_0$ become
$l=-k-m_0, l_0=-k-m$. Thus the ${\widehat P}_4$ degeneracy occurs when  both
$l,l_0$ are negative and satisfy $l< -k_0, l < k$.

Thus to cover all cases of both positive and negative values of the
parameter $l$ one needs to consider degeneracies induced by both 
${\widehat P}_2$ and ${\widehat P}_4$.

\section{The odd ${\widehat P}_1$ and ${\widehat P}_3$ automorphisms}
\label{section:oddref}
\subsection{The action of ${\widehat P}_1$  automorphism}
\label{subsection:P1}
Consider the two ${\widehat P}_1$ transformations
\begin{equation} G^k_m (x,\mathsf{a}) \stackrel{{\widehat P}_1}{\longrightarrow}
 G^m_k (-x,-a+2)=
1- F^k_m (x,\mathsf{a})\, ,
\label{sigma4G}
\end{equation}
and 
\begin{equation} F^k_m (x,\mathsf{a}) \stackrel{{\widehat P}_1}{\longrightarrow}
 F^m_k (-x,-a+2)=
1- G^k_m (x,\mathsf{a})\, ,
\label{sigma4F}
\end{equation}
where we took into consideration  that ${\widehat P}_1$ transforms
$x \to -x$, $a \to 2-a$ and $m \leftrightarrow k$ (which is equivalent to 
$\alpha_1 \leftrightarrow 
\alpha_4,$ , $\alpha_2 \leftrightarrow  \alpha_3$). Simultaneously, we have
${\widehat P}_1 ( F)=1-G$ and ${\widehat P}_1 (G)=1-F$ that we
inserted on the right hand sides.

Using identities \eqref{dualityRRxa} from subsection
\ref{subsection:gencase} we find for  $G^m_k (-x,-a+2)$
from \eqref{FGkm} using relation \eqref{dualityRRxa}:
\begin{equation}
\begin{split}
G^m_k (-x,-a+2)&= -2m 
\frac{R_{k(m+1)}^{(m)} (x,\mathsf{a}) R^{(m-2)}_{(k+1)(m-1)} (x,\mathsf{a}-2) }
{R_{m k}^{(m-1)} (x,\mathsf{a}) R_{(k+1)m}^{(m-1)}  (x,\mathsf{a}-2)}\\
&=1- F^k_m (x,\mathsf{a}) = 1-\frac{R_{m k}^{(m-1)} (x,\mathsf{a}-2) 
R_{m(k+1)}^{(m-1)} (x,\mathsf{a})}
{R_{m k}^{(m-1)} (x,\mathsf{a})
R_{m(k+1)}^{(m-1)}  (x,\mathsf{a}-2)}\, ,
\label{sigma4gf}
\end{split}
\end{equation}
or after multiplying both sides by $ R^{(m-1)}_{km} (x,\mathsf{a})/R_{m
(k+1)}^{(m-1)}  (x,\mathsf{a})$:
 \begin{equation}
\begin{split}
&\frac{R^{(m-2)}_{(m-1)(k+1)}(x, a-2) R^{(m)}_{(m+1) k} (x,\mathsf{a})}
{R_{m(k+1)}^{(m-1)} (x,\mathsf{a})R_{m(k+1)}^{(m-1)}  (x,\mathsf{a}-2)}\\
&=
\frac{1}{2m} \left(
\frac{ R^{(m-1)}_{mk} (x,\mathsf{a}-2) }
{ R_{m(k+1)}^{(m-1)}  (x,\mathsf{a}-2)}  %
-\frac{ R^{(m-1)}_{km} (x,\mathsf{a}) }
{ R_{m (k+1)}^{(m-1)}  (x,\mathsf{a})}\right) \, ,
 \end{split}
\label{sigma4gfa}
 \end{equation}
in which  we recognize
the master recurrence relation \eqref{masterrecurrence} !

Looking back on the remaining relation in equation
\eqref{sigma4F} it can be shown that it is equivalent to
recurrence relation
\eqref{dualmastera} dual to the master
recurrence relation \eqref{masterrecurrence}.

This completes the ${\widehat P}_1$ symmetry discussion showing how it nicely fits
the structure of solutions we have generated and fully characterized by
recursion relations. Thus the ${\widehat P}_1$ automorphism is
verified as long as the fundamental recursion relations hold and its
invariance can be considered a consistency check on the formalism.

\subsection{The ${\widehat P}_3$  automorphism}
\label{subsection:P3}
The automorphism ${\widehat P}_3$ transforms $x \to -x$ and ${\alpha}_1 
\leftrightarrow {\alpha}_2$ and  ${\alpha}_3 
\leftrightarrow {\alpha}_4$. For  $\alpha_i= 2 (l+k,-k,-m,1-l+m)$
it follows that ${\widehat P}_3(\alpha_i)=2(-k,l+k,1-l+m,-m)=
 2 ({\bar l}+{\bar k},-{\bar k},-{\bar m},1-{\bar l}+{\bar m})$ which 
 implies 
\[k\to {\bar k}=-l-k, \quad m \to {\bar m}=l-1-m\, .\]
Note that $\alpha_1+\alpha_2=2 l$ and $\alpha_3+\alpha_4=2(1- l)$ are
invariant under ${\widehat P}_3$  and so this time $l$ is invariant!  
Note that for ${\bar k}$ to be positive.
we must have $l<0$ with $\vert l \vert \ge k$. For ${\bar m}$ to be positive
we must have $l>0$ and $l \ge m+1$.
These are contradictory requirements and it looks that the ${\widehat P}_3$
transformed solution is not obtained by a simply transformation or 
relabeling of parameters.
However the transformation
\[{\widehat P}_3  :x \to -x , \; F \rightarrow G,\; G\rightarrow F, \; \alpha_1 \leftrightarrow 
\alpha_2,\; \alpha_3 \leftrightarrow \alpha_4 \, ,
\]
is realized as 
\[\begin{split}
&{\widehat P}_3 T_2^{-k} T_4^m \vert F=1, G=0\rangle_{(a,0,0,2-a)}=
 T_2^{k} T_4^{-m} \vert F=0, G=1\rangle_{(0,a,2-a,0)}\\
&={\widehat P}_3 (F_m^k (x,2 l), G_m^k (x,2l))
=(G_m^k (-x,2 l), F_m^k (-x,2 l)) \, .
\end{split}
\]
Substituting $k \to {\bar k}=-l-k$, $m \to {\bar m}=l-1-m$
with ${\bar l}=l$ into equation \begin{equation}\begin{split}
F_x &=F(F-1)(2G-1)-\frac{l+k-m}{x} F +
\frac{l+k}{x} \, ,\\
G_x &= -G (G-1)(2F-1) + \frac{l+k-m-1}{x} G
-\frac{k}{x} \, ,
\label{xhameqsmk}
\end{split}
\end{equation}
followed by 
$G \leftrightarrow F$ and $x\to -x$ transformation will leave equations 
\eqref{xhameqsmk} invariant verifying the invariance under
${\widehat P}_3$ automorphism. 

\section{Extending the method to $N>4$}
\label{section:higherN}
We consider few cases of solutions for higher $N=6$ and $N=8$ 
generated from specific seed solutions and illustrate the power of the
method based on automorphisms to determine the explicit form of solutions.
For convenience to describe the seed solution in such cases
we will be using the notation based on variables $j_i (z)$ of the even
dressing chain \eqref{dressingeqseven}.

\subsection{$N=6$ and $N=8$ and the seed solutions
 $j_i (z)=z/2, i=1,2,3$ and $j_i (z) = (-1)^{i+1} z/2, i=4,{\ldots},N$}
 \label{subsection:N68}
In this subsection we will consider solutions constructed
from seed solutions for even $N$ that of the type 
$j(z)=(z/2)(1,1,1,-1,1,-1)$
and $j(z)=(z/2)(1,1,1,-1,1,-1,1,-1)$.
Here we analyses such models for $N=6$ and $N=8$ but the
construction extends easily 
to all higher $N$.

We first consider a basic seed solution 
$j_i=\frac{z}{2} (1,1,1,-1,1,-1)$ with 
the  parameters $\alpha_i=(2-\mathsf{a}_2,\mathsf{a}_2, 0
,0,0,0)$ of $N=6$   dressing equations
 \eqref{dressingeqseven}. This seed solution solves
 the higher Painlev\'e equations \eqref{Nevengeneral}
with  $g_1=1, \; g_3=g_5=0,\; g_2=1,\;g_4=g_6=0$
and will be denoted by $\vert g_1=1,g_2=1\rangle_{(2-\mathsf{a},\mathsf{a}, 0
,0,0,0)}$.
For the above solution it holds that 
 $j_3+j_4=0, \, j_4+j_5=0, \,j_5+j_6=0, j_6+j_1=0$.
 As discussed in \cite{AGLZ2023}, to avoid division by zero 
 we need to exclude any action by 
$s_i, s_{i-1} \pi^{-1}, s_{i+1} \pi$ with $i=3,4,5,6$.
For similar reasons we also  need to exclude  the  translation operators
$T_i, T_{i+1}^{-1}, i=3,4,5,6 $. In conclusions we are led  to the 
only  finite solutions of the form:
\[
 T_1^{n_1} T_2^{n_2}  T_3^{-n_3}
 \vert g_1=1,g_2=1\rangle_{(2-\mathsf{a},\mathsf{a}, 0
,0,0,0)} ,\; n_1, n_3 \in \mathbb{Z}_+, \, n_2 \in \mathbb{Z}\, ,
 \]
with the parameters
\begin{equation}
(2-\mathsf{a}+2n_1-2n_2,\mathsf{a}+2 n_2+2n_3, -2 n_3
,0,0,0)\, .
\label{2-7orbits}
\end{equation}
One sees that the effect of $T_2$ is only to redefine  
$\mathsf{a}$ by a shift with $2 n_2$ as $T_1 $ did in the case of
the $N=4$ model.
According to the transformation rule 
\eqref{Rntranslation} we find that  the above three translation operators
are being connected via the automorphism ${\widehat P}_3$ :
\begin{equation} {\widehat P}_3  T_3^{-1} {\widehat P}_3= T_{1},\; \;
	{\widehat P}_3 T_2 {\widehat P}_3= T_{2}^{-1}
\label{p3T3T2}
\end{equation}
We can now investigate the recurrence relations emerging  on the orbit of
$T_3^{-1}$ to express it in terms of Kummer polynomials and build the
complete solutions utilizing fully the ${\widehat P}_3$ automorphism 
to ensure invariance under:
$x \to-x,a \to 2-a, n_2 \to n_4, n_4 \to n_2$.

Extending the solution to $N=8$ using analogous  configuration of 
$
j_i (z)=z/2, i=1,2,3$ and $j_i (z) = (-1)^{i+1} z/2, i=4,{\ldots},8$
we can now repeat the above analysis
to arrive at the same form of solutions
\[
 T_1^{n_1} T_2^{n_2}  T_3^{-n_3}
 \vert g_1=1,g_2=1\rangle_{(2-\mathsf{a},\mathsf{a}, 0
,0,0,0,0,0)} ,\; n_1, n_3 \in \mathbb{Z}_+, \, n_2 \in \mathbb{Z}\, ,
 \]
with the parameters
\begin{equation}
(2-\mathsf{a}+2n_1-2n_2,\mathsf{a}+2 n_2+2n_3, -2 n_3
,0,0,0,0,0)\, .
\label{2-8orbits}
\end{equation}
The translation operators will continue to be connected via the
automorphism ${\widehat P}_3$ as in relation \eqref{p3T3T2}.

\subsection{$N=6$ and the seed solutions $j(z)=(z/2)(1,1,-1,1,1,-1)$}
\label{subsection:N6twoneg}
As shown in \cite{AGLZ2023} the configuration :
\begin{equation}j_i=\frac{z}{2} (1,1,-1,1,1,-1)
,\; \quad \alpha_i=(2-\mathsf{a}_4, 0,0,\mathsf{a}_4,0,0)\, ,
\label{oneminusone}
\end{equation}
is another example of the seed solutions of $N=6$  dressing chain
equations \eqref{dressingeqseven} with two negative components of $j_i(z)$
that this time have two positive components between them.

As shown in \cite{AGLZ2023} the general rational solutions are then  
obtained by acting  with 
 \[
 T_1^{n_1} T_4^{n_4}  T_2^{-n_2} T_5^{-n_5}, n_1, n_2,n_4,n_5  \in \mathbb{Z}_+
 \, ,\]
 on the  seed solution \eqref{oneminusone} with the
parameters: 
\begin{equation}
(2-\mathsf{a}_4+2n_1+2n_2, -2n_2,-2n_4,\mathsf{a}_4+2n_4+2n_5,-2n_5,-2n_1)\,.
\label{2-8orbits8}
\end{equation}
According to the transformation rule 
\eqref{Rntranslation} we find that  the 
automorphisms ${\widehat P}_5$ and ${\widehat P}_2$ will connect the above translation
operators as follows:
\begin{equation}
\begin{split}
{\widehat P}_5  T_1 {\widehat P}_5&= T_{5}^{-1},\; \;
	{\widehat P}_5 T_4 {\widehat P}_5= T_{2}^{-1}\\
{\widehat P}_2  T_1 {\widehat P}_2&= T_{2}^{-1},\; \;
	{\widehat P}_2 T_4 {\widehat P}_2= T_{5}^{-1}
	\end{split}
\label{p5T1T5T4T2}
\end{equation}
  This should allow us to construct general solutions out of Kummer-like
  polynomials.
  \section{Summary and conclusions}
  The even periodicity higher Painlev\'e equations possess extended  symmetry structure 
  when  compared with the odd periodicity counterparts. The additional reflection automorphisms that
  we have introduced in this work split on reflection around even and odd points and subsequently  play different roles
  in determining rational solutions. The reflections around even points generate new solutions while maintaining the  parameters of Painlev\'e equations unchanged thus they are directly engaged in generating degeneracy. The reflections around odd points act as duality transformations that maintain the value of solutions although formally changing its appearance by mapping into the dual formulation. The duality property plays crucial role in determining the structure of solutions and provide a useful framework for merging solutions of simple orbits of one single translation operator into orbits created by actions of two independent translation operators. Developing these formalism embed higher number of translation operators will be crucial for complete characterization of more complex solutions including Umemura solutions, We palne to complete proofs for N byond N =4. It would be important to establish theory of discrete Painleve structures  that are naturally e emerging from use of translation operators
  Proofs given for $N=4$ and arguments presented by looking for particular solutions will be established for higher N  >4.

\appendix
\section{Matrix and braids representation of reflection automorphisms
for $N=4$}
\label{appendix:matrices}

It is illuminating to  describe the reflection automorphisms
in the language of braids and matrices.  We give our discussion for
$N=4$ but it can easily be generalized to higher $N$.

We start with braids and separate 
automorphisms  into two classes. 
The first class consists of 
those automorphisms that  can be graphically represented by 
 braids assigned to a 
single  segment  with upper/lower boundaries
consisting of four points (1,2,3,4). 
This layout allows for crossing of two lines ($13$ for $P_4$ and $24$
for $P_2$) while the two remaining lines do not transpose as shown on the next two
figures.
			
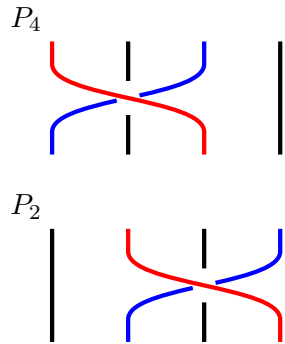
\begin{center}
		\begin{tikzpicture}
			\pic[	
			braid/.cd,
			line width=8pt,
			number of strands=4,
			ultra thick,
			strand 1/.style={red},
			strand 2/.style={black},
			strand 3/.style={blue},
			strand 4/.style={black},
			] {braid={s_{1,3}}};
			\node[above left] {$P_4$};
		\end{tikzpicture}
		\end{center}
				\begin{center}
			\begin{tikzpicture}
				\pic[	
				braid/.cd,
				line width=8pt,
				number of strands=4,
				ultra thick,
				strand 1/.style={black},
				strand 2/.style={red},
				strand 3/.style={black},
				strand 4/.style={blue},
				] {braid={s_{2,4}}};
				\node[above left] {$P_2$};
			\end{tikzpicture}
		\end{center}
The next class of automorphisms  $P_3$ and $P_1$
act graphically over two segments by transposing lines
with each 
transposition being allowed only one  single crossing of lines in every 
segment. 
		
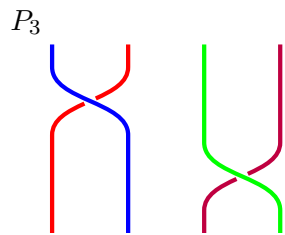
\begin{center}
			\begin{tikzpicture}
				\pic[	
				braid/.cd,
				line width=8pt,
				number of strands=4,
				ultra thick,
				strand 1/.style={blue},
				strand 2/.style={red},
				strand 3/.style={green},
				strand 4/.style={purple},
				] {braid={s_{1,2}s_{3,4}}};
				\node[above left] {
				$P_3$};
			\end{tikzpicture}
		\end{center}
		
		
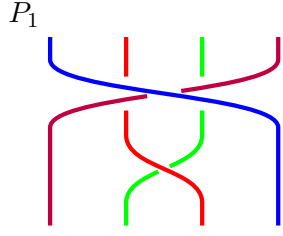
\begin{center}
			\begin{tikzpicture}
				\pic[	
				braid/.cd,
				line width=8pt,
				number of strands=4,
				ultra thick,
				strand 1/.style={blue},
				strand 2/.style={red},
				strand 3/.style={green},
				strand 4/.style={purple},
				] {braid={s_{1,4}s_{2,3}}};
				\node[above left] {$P_1$};
			\end{tikzpicture}
		\end{center}
Automorphisms $P_{2i+1}, i=0,1$ transpose the line $1$ into the
even lines $4$ and $2$. If these lines cross in the first segment,
then the crossings of the remaining lines $23$ and $41$ occurs in the
second segment.

Next, let us %
represent the reflection automorphisms 
as matrices acting on columns consisting  of points $1,2,3,4$. All
matrix elements will be either $0$ or $1$.
The even reflections will have two diagonal elements equal to $1$
corresponding to the two fixed points:
\[
P_2 :   \begin{pmatrix} 1 &0 &0&0\\
0 &0 &0&1\\
0 &0 &1&0\\
 0&1 &0 &0
 \end{pmatrix} \begin{pmatrix} 1\\2\\3\\4  \end{pmatrix} 
 = \begin{pmatrix} 1\\4\\3\\2  \end{pmatrix} 
\]
\[
P_4 :   \begin{pmatrix} 0 &0 &1&0\\
0 &1 &0&0\\
1 &0 &0&0\\
 0&0 &0 &1
 \end{pmatrix} \begin{pmatrix} 1\\2\\3\\4  \end{pmatrix} 
 = \begin{pmatrix} 3\\2\\1\\4  \end{pmatrix} 
\]
Thus $Tr (P_2)=Tr (P_4)=2$. The odd reflections are traceless: $Tr
(P_1)=Tr (P_3)=0$:
\[
P_1 :   \begin{pmatrix} 0 &0 &0&1\\
0 &0 &1&0\\
0 &1 &0&0\\
 1&0 &0 &0
 \end{pmatrix} \begin{pmatrix} 1\\2\\3\\4  \end{pmatrix} 
 = \begin{pmatrix} 4\\3\\2\\1  \end{pmatrix} 
\]
\[
P_3 :   \begin{pmatrix} 0 &1 &0&0\\
1 &0 &0&0\\
0 &0 &0&1\\
 0&0 &1 &0
 \end{pmatrix} \begin{pmatrix} 1\\2\\3\\4  \end{pmatrix} 
 = \begin{pmatrix} 2\\1\\4\\3  \end{pmatrix} 
\]
The determinants are $\vert P_i \vert=-(-1)^i$ with even matrices 
having one eigenvalue $-1$ and the remaining three eigenvalues $+1$,
the odd matrices have two eigenvalues equal to $-1$ and two equal to $+1$.
The matrices are all
symmetric $ P_i^T= P_i$ and orthogonal. All rows are orthogonal
and $P_i^2=1$.

The even matrices commute and the odd matrices commute among themselves. Also
\[ 
P_2 P_4=P_1P_3= P_4P_2=P_3P_1=P_1P_2P_3P_4=
   \begin{pmatrix} 0 &0 &1&0\\
0 &0 &0&1\\
1 &0 &0&0\\
 0&1 &0 &0
 \end{pmatrix} =\pi^{2}=\pi^{-2}
\]
and 
\[P_1P_2=P_2P_3=P_3P_4=P_4P_1
= \begin{pmatrix} 0 &1 &0&0\\
0 &0 &1&0\\
0 &0 &0&1\\
1&0 &0 &0
 \end{pmatrix} =\pi^{-1}
\]
And
\[P_4P_3=P_3P_2=P_2P_1=P_1P_4
= \begin{pmatrix} 0 &0 &0&1\\
1 &0 &0&0\\
0 &1 &0&0\\
0&0 &1 &0
 \end{pmatrix} =\pi
\]
Eigenvalues of $\pi$ and $\pi^{-1}$ are $-1,+1,- I, +I$ so the
determinant is $-1$. Eigenvalues of $\pi^2$ are $-1,1,-1,1$ so that the
determinant is $+1$. All $\pi^i$ are traceless.

It is easy to verify explicitly that reflections together with
automorphisms $\pi^n$ close under multiplications and therefore form a 
a group $\{ P_{n_1}, \pi^{n_2} \}$ with the identity matrix $1$. For
example, we find easily in the above matrix representation that
\[ P_n \pi = P_{n-1}\]
and that the product $P_2 P_4P_1P_3$ is
equal to $1$ as found by general arguments in the text.

\end{document}